\pdfoutput=1
\newif\ifarxiv
\arxivtrue
\newif\iffull
\fulltrue
\newif\ifanonymized
\anonymizedfalse

\ifarxiv
\documentclass[acmsmall,authorversion,nonacm]{acmart}
\else
\documentclass[acmsmall,review]{acmart}
\fi

\usepackage{graphicx}
\usepackage{xcolor}
\usepackage{wrapfig}
\usepackage{comment}
\usepackage{xspace}
\usepackage[frozencache,cachedir=.]{minted}
\usepackage{url}
\usepackage{tcolorbox}
\usepackage{alltt}
\usepackage{amsmath}
\usepackage{amsfonts}
\usepackage{natbib}

\ifarxiv
\setcopyright{none}
\copyrightyear{2021}
\acmYear{2021}
\acmDOI{10.1145/1122445.1122456}

\acmConference{None}
\acmBooktitle{None}
\acmVolume{1}
\acmNumber{1}
\acmArticle{1}
\acmMonth{1}

\else
\setcopyright{acmcopyright}
\copyrightyear{2021}
\acmYear{2021}
\acmDOI{10.1145/1122445.1122456}

\acmJournal{JACM}
\acmVolume{37}
\acmNumber{4}
\acmArticle{111}
\acmMonth{8}
\fi

\newcommand{\fstar}{F$^\star$\xspace}
\newcommand{\MPT}{MPT\xspace}

\newcommand\model[1]{[\![{#1}]\!]}
\newcommand{\fstarinline}[1]{\mintinline[escapeinside=!!]{OCaml}{#1}\xspace}

\newcommand{\emphmarker}[1]{\colorbox[rgb]{0.8,1.0,0.8}{#1}}

\newcommand\DOM{\mathbf{Dom}}

\DeclareMathSymbol{\mlq}{\mathord}{operators}{``}
\DeclareMathSymbol{\mrq}{\mathord}{operators}{`'}

\newcommand{\mathfstar}[1]{\mathrm{\fstarinline{#1}}}

\newif\ifdraftComments
\draftCommentsfalse
\def\mkDraftFn#1#2{%
  \expandafter\def\csname #1\endcsname##1{\ifdraftComments\textcolor{#2}{[#1: ##1]}\marginpar[$\longrightarrow$]{$\longleftarrow$}\fi}%
}
\mkDraftFn{satos}{red}
\mkDraftFn{KS}{green}
\mkDraftFn{AI}{magenta}
\mkDraftFn{JF}{blue}

\newtheorem{theorem}{Theorem}

\newtheorem{assumption}{Assumption}
 
\newtheorem{codeexample}{Example}

\newtheorem{remark}{Remark}

\begin{document}

\title[Verification of a Merkle Patricia Tree Library Using \fstar]{Verification of a Merkle Patricia Tree Library Using \fstar}
% \satos{If we add \texttt{\thanks{supported by}} to the title, the compiler fails}]


\ifanonymized
\author{Anonymized}
\else

\author{Sota Sato}
\orcid{0000-0003-1648-1841}
\affiliation{%
  \institution{Kyoto University}
  \country{Japan}}

\author{Ryotaro Banno}
\orcid{0000-0003-1648-1841}
\affiliation{%
  \institution{Kyoto University}
  \country{Japan}}

\author{Jun Furuse}
\affiliation{%
  \institution{Dailambda Inc.}
  \country{Japan}}

\author{Kohei Suenaga}
\orcid{0000-0002-7466-8789}
\affiliation{%
  \institution{Kyoto University}
  \country{Japan}}

\author{Atsushi Igarashi}
\orcid{0000-0002-5143-9764}
\affiliation{%
  \institution{Kyoto University}
  \country{Japan}}

\fi
%


\newcommand{\colorR}[1]{\textcolor{red}{#1}}
\newcommand{\pagelimitmarker}[1]{~\\ {\colorR{\ifthenelse{\thepage>#1}{\Huge Exceeding the page limit}{\huge Within the page limit}}}~\\ {\huge{\colorR{~~Page Limit\,\,\,\,\, = #1}}}~\\ {\huge{\colorR{~~Current Page = $\thepage$}}}}

\ifanonymized
	\renewcommand{\shortauthors}{Anonymized}
\else
	\renewcommand{\shortauthors}{Trovato and Tobin, et al.}
\fi

\begin{abstract}

  A \emph{Merkle tree} is a data structure for representing a key--value store as a tree.
  Each node of a Merkle tree is equipped with a hash value computed from those of their descendants.
  A Merkle tree is often used for representing a state of a blockchain system such as Ethereum and Tezos since it can be used for efficiently auditing the state in a trustless manner.
  % and calculating the hash value of the store, with which 
  Due to the safety-critical nature of blockchains, ensuring the correctness of their implementation is paramount.

  We show our formally verified implementation of the core part of \emph{Plebeia} using \fstar, a programming language to implement a formally verified functional program.
  Plebeia, which is implemented in OCaml, is a library to manipulate an extension of Merkle trees (called \emph{Plebeia trees}).
  It is being implemented as a part of the storage system of the Tezos blockchain system.
  To this end, we \emph{gradually} ported Plebeia to \fstar; the OCaml code extracted from the modules ported to \fstar is linked with the unverified part of Plebeia.
  By this gradual porting process, we can obtain a working code from our partially verified implementation of Plebeia; we confirmed that the binary passes all the unit tests of Plebeia.
  % continuously obtain a partially-verified working implementation before porting the entire code of Plebeia.

  %
  More specifically, we verified the following properties on the implementation of Plebeia:
  (1) Each tree-manipulating function preserves the invariants on the data structure of a Plebeia tree and satisfies the functional requirements as a nested key--value store;
  (2) Each function for serializing/deserializing a Plebeia tree to/from the low-level storage is implemented correctly; and
  (3) The hash function for a Plebeia tree is relatively collision-resistant with respect to the cryptographic safety of the blake2b hash function.
  During porting Plebeia to \fstar, we found a bug in an old version of Plebeia, which was overlooked by the tests bundled with the original implementation.
  To the best of our knowledge, this is the first work that verifies a production-level implementation of a Merkle-tree library by \fstar.

\end{abstract}

% \input{ccs.tex}


\maketitle

\section{Introduction}

Blockchain-based cryptocurrencies, such as Bitcoin~\cite{nakamotobitcoin} and
Ethereum~\cite{wood2014ethereum}, have become an important infrastructure to
exchange virtual assets securely yet without a central authority.
Any vulnerabilities~\cite{cve1,cve2} are potential attack vectors of malicious hackers because
they are databases carrying a huge amount of financial values.
Indeed, such attacks have caused significant financial losses in the past~\cite{daoattack}.
Therefore, formally verifying the security of these frameworks is becoming a hot topic
in PL research~\cite{vyper,2017obsidian,bhargavan2016formal}.
% \satos{We have to specify correct date instead of n.d.?(What is the date of documentation?)}

% \KS{This paragraph may be misleading.}
% The notion of ``the security of a blockchain system''  is diverse, as exemplified by the known attacks against blockchain systems~\cite{cve1,cve2}.
% Some notorious attacks, such as the DAO attack~\cite{daoattack}, target the vulnerabilities in smart contracts, which
% are computer programs that are registered to a blockchain for automated transactions;
% there are other known attacks that exploit vulnerabilities of the protocols,
% such as taking over the majority of the share to control the blockchain.

In this context, the integrity of blockchain states, including transaction records and the (persistent) states of smart contracts,
is another fundamental security property.  A popular technique to ensure
the integrity of blockchain states is to implement it using a data structure called
\emph{Merkle trees}~\cite{merkle1989certified},
which are also called hash trees, and their extensions.  Roughly, a Merkle tree is a tree whose leaves store data
and whose internal nodes store the hash value computed from their children.  Merkle trees
enable efficient and trustless verification of the existence of a piece of information in a tree.  Ethereum uses Merkle Patricia Trees (MPTs),
which are a combination of Merkle trees and Patricia trees, to
implement secure key--value stores.

Since time and space efficiency is important for the user experience of a blockchain system,
various improvements of Merkle trees are being intensively developed.
In Tezos~\cite{goodman2014tezos}, one of the blockchain-based cryptocurrencies, this effort led to the storage system called \emph{Plebeia}.
Plebeia implements a variant of \MPT{}s---called \emph{Plebeia trees}---that is highly optimized to reduce the disk storage
size\footnote{\url{https://www.dailambda.jp/blog/2019-08-08-plebeia/}}.
% Plebeia uses techniques such as Huet's zippers~\cite{huet1997zipper} to reduce its execution time.  
Although each technique used in Plebeia is well known, the correctness of the entire implementation, which combines them
in a sophisticated way, is nontrivial.

In this paper, we present our formally verified implementation of the core part of Plebeia.
We verified the following properties of the current Plebeia implementation:
  (1) Each tree-manipulating function preserves the invariants on the data structure of a Plebeia tree and satisfies the functional requirements as a nested key--value store;
  (2) Each function for serializing/deserializing a Plebeia tree to/from the low-level storage is implemented correctly;
  (3) The hash function for a Plebeia tree is relatively collision-resistant with respect to the cryptographic safety of the blake2b hash function.
% In addition, we verified (3) the hash function defined on the Merkle tree is relatively collision resistant assuming
% the cryptographic safety of the blake2b hash function.
To this end, we ported the core part of the Plebeia, which is implemented in OCaml, to \fstar~\cite{mumon},
which is a programming language for formal verification,
and wrote the proofs of these properties.
The \fstar type checker automatically verifies that our proofs are correct.

Since we ported only a part of Plebeia to \fstar, our \fstar code cannot be built to working binary by itself.
To address this problem, we use the code-extraction functionality of the \fstar compiler that extracts OCaml code from machine-checked proofs.
Using this functionality, we extracted OCaml code from our proof and linked it with the unverified part of Plebeia.
We confirmed that the resulting code runs and successfully passes all the tests bundled with Plebeia.

% By this gradual porting process, we can continuously obtain a partially-verified working implementation before porting the entire code of Plebeia.

% \AI{We may want to reveal that we don't even deal with invariants related to hashes at some point (but where)?}
% Although Plebeia trees involve hashes for ensuring integrity, we do not deal with
% hash safety and cryptographic correctness properties.
% Rather, our main interest in this paper is in the correctness of tree manipulations,
% complicated by the combination of Patricia trees and the zippers.

The rest of this paper is organized as follows.
Section~\ref{chap:fstar} reviews \fstar;
Section~\ref{chap:plebeiatrees} introduces Plebeia trees and the structural and functional requirements of a Plebeia tree to be satisfied;
Section~\ref{chap:extension} explains the extensions to Plebeia trees implemented in the actual code;
Section~\ref{chap:verification} explains how we verified the core part of Plebeia with \fstar.
After the discussion on related work in Section~\ref{chap:relatedwork},
Section~\ref{chap:conclusion} concludes.
% \satos{section reference should be fixed (or can it be fixed by setting?)}

\ifanonymized
All the \fstar source code is available in the supplementary material.
\else
All the \fstar source code is available at \url{https://gitlab.com/dailambda/plebeia/-/tree/banno@port_into_fstar/fstar}.
\fi
\section{\fstar}
\label{chap:fstar}

\fstar~\cite{mumon} is a functional programming language for program verification.
By using \fstar, one can formalize the properties of a program and prove it.
This section briefly introduces the features of \fstar, which we used to verify the core part of Plebeia.
For a detailed exposition, see \url{http://www.fstar-lang.org/tutorial/}.

\subsection{Syntax}
\label{fstar:syntax}

The syntax of \fstar is similar to that of OCaml, as the following example shows.
\begin{codeexample}
  The following is the definition of a function \fstarinline{sum} that takes an integer list and returns the sum of the integers in the list.
  \begin{minted}{OCaml}
  val sum : list int -> int
  let rec sum = function | [] -> 0 | x :: xs -> sum xs + x
  \end{minted}
  \label{fstar:sum}
\end{codeexample}

We remark the following differences of the \fstar syntax from the OCaml syntax:
\begin{itemize}
  \item  In \fstar, a function definition can be preceded by a type declaration, whereas a type declaration has to be written in a separate signature definition in OCaml.
  \item Type parameters of a polymorphic type in \fstar are written like \fstarinline{t 'a 'b, ...}, whereas it is written \fstarinline{('a,'b,...) t} in OCaml.
  \end{itemize}

\subsection{Effects}
\label{fstar:effect-system}

The syntax of a function type in \fstar can be optionally accompanied by
an \emph{effect}; the syntax of a function type is
$T_1 \texttt{->} \mathit{Eff}\ T_2$, where $T_1$ is the type of the argument,
$T_2$ is the type of the return value, and $\mathit{Eff}$ is an effect.
An effect $\mathit{Eff}$ overapproximates the side effects 
that the function may cause when it is called.
\fstar provides several primitive effects, including \fstarinline{Tot},
which indicates that functions do not cause any side effects and it terminates for any arguments.
\fstarinline{ST} indicates that functions may read from and write to memory.

The effect of a function is treated as \fstarinline{Tot} if it is omitted.
Hereafter, we say that a function is \emph{pure}  if the function has the effect \fstarinline{Tot}.

\subsection{Refinement Types}

The type system of \fstar incorporates \emph{refinement types}~\cite{DBLP:conf/pldi/FreemanP91}.
A refinement type is a type constrained by a logical predicate.
It can express a property of values that is more detailed than one expressed by a simple type.

A binding of a variable to a refinement type is written \fstarinline{x:T{P}}
where \fstarinline{T} is a type and \fstarinline{P} is a predicate;
this refinement type is for values that have type \fstarinline{T} and satisfy 
the predicate \fstarinline{P}. For example, \fstarinline{x:int{x > 0}}
is a binding of \fstarinline{x} to the type for positive integers.

A programmer can use
(1) comparison operators (e.g., \fstarinline{=}, \fstarinline{>=});
(2) logical operators (e.g., \verb|/\|, \verb|\/|, \fstarinline{forall}, \fstarinline{exists}); and
(3) user-defined pure functions that return a value of type \fstarinline{bool}.
For example, the following type declaration of \fstarinline{sum} expresses that
it only accepts a list of positive integers and it returns a positive integer.
\begin{minted}{OCaml}
  val sum : list (p:int{p > 0}) -> r:int{r > 0}
\end{minted}

One can use a \emph{dependent function type} to express a relation between the arguments and the return value of a function.
For example, the type of function \fstarinline{sum} can be made more precise by the following (dependent) function type.
\begin{minted}{OCaml}
  val sum : (v:list (p:int{p > 0})) -> (r:int{r >= List.length v})
\end{minted}
This type declaration asserts that the return value \fstarinline{r} of 
the function \fstarinline{sum} is greater than or equals to the length of the argument list
\fstarinline{v}.

We use another notation that \fstar provides, which the following example illustrates, to express a type of a function with the effect \fstarinline{ST}.
\begin{minted}[escapeinside=!!]{OCaml}
  val f : x:!$\tau_1$! -> ST !$\tau_2$! (requires (fun h -> !$P_1$!)) (ensures (fun h0 x h1 -> !$P_2$!))
\end{minted}
The function of the above type takes \fstarinline{x} of type $\tau_1$ and returns a value of type $\tau_2$.
The effect \fstarinline{ST} indicates that this function may read/write from/to memory.
The clauses \fstarinline{requires} and \fstarinline{ensures} indicate the precondition and the postcondition of this function.
The clause \fstarinline{(requires (fun h -> !$P_1$!))}, where \fstarinline{h} is of an abstract type representing heap and $P_1$ is a predicate, expresses that \fstarinline{f} is supposed to be called where heap \fstarinline{h} satisfies $P_1$.
The clause \fstarinline{(ensures (fun h0 x h1 -> !$P_2$!))}, where $P_2$ is a predicate, expresses that the call to \fstarinline{f} modifies heap as described by $P_2$, where \fstarinline{h0} is the heap before the call, \fstarinline{h1} is the heap after the call, and \fstarinline{x} is the value of type $\tau_2$ returned by this function.
By using the predicate $P_2$, one can express the relation among the returned value, the heap before the call, and one after the call.

\fstar tries to automatically typecheck a definition against a refinement type
by discharging the verification condition with an SMT solver.
The SMT solver Z3~\cite{z3solver} is used as an SMT solver in our verification.

The type checker of \fstar has to prove the termination of a function if it is given the effect \fstarinline{Tot},
which is in general undecidable.
The \fstar type checker verifies the termination according to the following strategy.
\begin{itemize}
  \item By default, \fstar tries to prove that the argument of every recursive call
  in a function definition is structurally decreasing in a certain well-founded measure
  specified for the type of the argument.  This heuristic is sufficient to prove
  the termination of the function \fstarinline{sum} in Example~\ref{fstar:sum} automatically
  because the length of the list passed as the first argument of every recursive call to \fstarinline{sum}
  is shorter than that of the original list.  
  
  \begin{sloppypar}
  \item One can supply a hint on which argument is structurally decreasing; 
  then \fstar tries to prove that this argument is indeed decreasing.
  For example, the termination of the following \texttt{sum'}, 
  which is a variant of \texttt{sum} with an accumulator argument,
  cannot be automatically proved by \fstar because its first argument is not decreasing.
  \end{sloppypar}
  \begin{minted}{OCaml}
    val sum' : int -> list int -> Tot int
    let rec sum' c = function | [] -> c | x :: xs -> sum' (x + c) xs
  \end{minted}
  However, \fstar can prove the termination after giving a hint by changing the type-declaration part to the following one
  whose \fstarinline{(decreases v)} annotation indicates that, in every recursive call,
  the length of the second argument is strictly less than that of \fstarinline{v}.
  \begin{minted}{OCaml}
    val sum' : int -> (v:list int) -> Tot int (decreases v)
  \end{minted}
  \fstar generates the verification conditions for the length of the second argument is indeed decreasing and
  tries to discharge the condition by Z3.  For this example of \fstarinline{sum'},
  Z3 successfully discharges these conditions.
\end{itemize}

\subsection{A Theorem as a Dependently-Typed Function}

A dependent function type is used to state a theorem in \fstar.
For example, the following type declaration states and proves a theorem
\fstarinline{n_plus_0} that asserts $\forall n \in \mathbb{Z}. n + 0 = n$.
\begin{minted}{OCaml}
  val n_plus_0 : (n:int) -> (_:unit{n + 0 = n})
  let n_plus_0 n = ()
\end{minted}
A theorem is, as the above example shows, expressed as a dependent function type
whose return type is a refinement of \fstarinline{unit}.
The proof of a theorem (stated as a type declaration) is given by a function definition of this type.
In fact, for the above \fstarinline{n_plus_0}, it suffices to give \fstarinline{()} as its body
because Z3 SMT solver can automatically discharge $\forall n \in \mathbb{Z}. n + 0 = n$.

Another way of stating a theorem in \fstar is using a notation \fstarinline{Lemma (requires !$P$!) (ensures !$Q$!)}, where $P$ and $Q$ are predicates.
For example, one can state the theorem $\forall n \in \mathbb{Z}. n + 0 = n$ as follows.
\begin{minted}{OCaml}
  val n_plus_0 : (n:int) -> Lemma (requires True) (ensures n + 0 = n)
\end{minted}
If $P$ is \fstarinline{True}, then we can omit the \fstarinline{requires} clause and the \fstarinline{ensures} clause.
\begin{minted}{OCaml}
  val n_plus_0 : (n:int) -> Lemma (n + 0 = n)
\end{minted}
In this paper, a theorem stated using the notation \fstarinline{Lemma (requires !$P$!) (ensures !$Q$!)} can be considered equivalent to the one mentioned above that stated using a refinement of \fstarinline{unit}.  See~\citet{DBLP:conf/cpp/GrimmMFHMPRRSB18} for detail.

\begin{sloppypar}
We can prove a theorem that requires to use induction in \fstar.
For example, consider the equality \fstarinline{c + sum v = sum' v c}.
where \fstarinline{sum} and \fstarinline{sum'} are those defined above.
We can state that the above equality holds for any \fstarinline{c} and \fstarinline{v} in \fstar as follows:
\end{sloppypar}
\begin{minted}{OCaml}
val sum_theorem : (v:list int) -> (c:int) -> Lemma (c + sum v = sum' c v)
\end{minted}
This theorem can be proved by mathematical induction on the length of the list \fstarinline{v},
which is expressed as the following function definition.
\begin{minted}{OCaml}
let rec sum_theorem v c = match v with | [] -> ()
  | x :: w -> sum_theorem w (x + c); ()
\end{minted}
\begin{sloppypar}
The pattern-matching clause for \fstarinline{[]} corresponds to the base case of the induction, 
and the pattern-matching clause for \fstarinline{x::w} corresponds to the step case.
In the base case, since the list \fstarinline{v} is \fstarinline{[]},
the equation is reduced to \fstarinline{c + 0 = c}.
Since the reduced equation is simple enough to be discharged by Z3 SMT solver,
a programmer does not have to write anything other than \fstarinline{()} as a proof.
In the step case, the list \fstarinline{v} can be decomposed to 
an integer \fstarinline{x} and the remaining list \fstarinline{w}.
The equation is reduced to \fstarinline{c + x + sum w = sum' (c + x) w}.
To derive this equation, we use the induction hypothesis
\fstarinline{sum_theorem w (x + c)}, which is expressed by the call \fstarinline{sum_theorem w (x+c)} in the above example.
\fstar adds this fact to the list of available assumptions at the point where \fstarinline{sum_theorem w (x + c)} is called.
\end{sloppypar}

\section{Plebeia Trees}
\label{chap:plebeiatrees}

A \emph{Plebeia tree} is an extension of a Merkle Patricia tree (MPT), which is a tree whose node keeps a hash value computed by those of its descendants.
To explain a Plebeia tree and the invariants imposed on its structure, this section first explains an MPT.
We will explain the most basic definition of Plebeia trees in this section.
The implementation detail, which comes with various extensions, is explained in Section~\ref{chap:extension}.

\subsection{Merkle Patricia Tree}
\label{sec:mpt}

A Merkle Patricia tree (MPT) is a Patricia tree~\cite{patriciatree} (a binary radix tree) whose nodes are associated with hashes.
An MPT is defined by the following \fstar type.
\begin{minted}[escapeinside=!!]{OCaml}
type side = L | R and key = list side and hash = string
(* The type for MPTs *)
type node = | Empty
            | Leaf of !\texttt{value}! * hash  (* !\texttt{value}! here is the type of values *)
            | Internal of key * node * key * node * hash
\end{minted}
The type \fstarinline{key} represents non-empty lists of bits \fstarinline{L} and \fstarinline{R}.
A key is associated with each edge in an MPT.
Each node is accompanied by a \fstarinline{hash} represented by a string.

\begin{sloppypar}
A node of MPT is either an empty tree (\fstarinline{Empty}), a leaf \fstarinline{Leaf(v,h)}, or
an internal node \fstarinline{Internal(k1,n1,k2,n2,h)}.
\fstarinline{Leaf(v,h)} has a value \fstarinline{v} and is associated with its hash \fstarinline{h}.
\fstarinline{Internal(k1,n1,k2,n2,h)} is the parent of the left child \fstarinline{n1} and the right child \fstarinline{n2},
and is associated with its hash \fstarinline{h}.
The key \fstarinline{k1} is on the edge from the internal node to \fstarinline{n1};
the key \fstarinline{k2} is on the edge to \fstarinline{n2}.  
\end{sloppypar}

A \emph{Merkle hash}~\cite{merkle1989certified} is a hash of an MPT node that is computed in the following way:
\begin{itemize}
\item The Merkle hash of a leaf is the hash of its value.
\item The Merkle hash of an internal node is computed by hashing the tuple of (1) the Merkle hashes of its subnodes and (2) the edge labels.
\end{itemize}
Practically assuming that there is no hash collision, the Merkle hash of the root node can identify an MPT:
two trees are identical if and only if their Merkle hashes are the same under the no-hash-collision assumption.
This property is suitable for blockchains;
for example, participants in a blockchain network can efficiently verify
that they share the same blockchain states by exchanging their Merkle hashes of the root node
instead of communicating an (often large) MPT itself.

\subsection{Plebeia Tree}
\label{sec:typeForPlebeiaTrees}

\subsubsection{Motivation and Definition}
One of the major problems in a na\"{\i}vely implemented MPT is the internal fragmentation of storage
that is organized as a set of fixed-size records.  To observe the problem,
suppose that (1) the storage size required to keep a value or a pointer is 32 bits,
(2) the size of a hash is 256 bits, and (3) the size of a key is $k$ bits on average.
Then, the size to store a leaf is $32 + 256 = 288$ bits, whereas the size for an internal node is
$k + 32 + k + 32 + 256 = 2k + 320$ bits.  If $k$ is close to $200$,
which is often the case in the current Tezos blockchain, the size for an internal node is
2.5 times as large as that for a leaf.
This imbalance in the sizes of a leaf and an internal node leads to wasted space in each fixed-size record in storage.
To address this issue, Plebeia defines two kinds of internal nodes: a \emph{branch} and an \emph{extender}.
Plebeia also has another kind of nodes called \emph{buds}, .

This extended MPT is called a \emph{Plebeia tree}, whose  type definition is as follows:
\begin{minted}[escapeinside=!!]{OCaml}
  type node = | Leaf of !\texttt{value}! * hash
              | Branch of node * node * hash
              | Extender of key * node
\end{minted}
\noindent
A node \fstarinline{Branch(n1,n2,h)} has a Merkle hash \fstarinline{h} and two subnodes \fstarinline{n1} and \fstarinline{n2},
which are connected with edges with one-bit keys \fstarinline{L} and \fstarinline{R} (therefore omitted).
A node \fstarinline{Extender(k,n)} has one subnode \fstarinline{n};
the edge to \fstarinline{n} has a non-empty key \fstarinline{k}.
The hash field of a node \fstarinline{Extender(k,n)} is omitted; its
hash value is defined as the concatenation of the hash value of its child node \fstarinline{n} and its key \fstarinline{k}, and hence its hash value can be calculated without hashing.
Notice that an extender shoulders the role of bringing a key on the edges to its subnodes,
which is the responsibility of an internal node in an MPT.
Actually, this type cannot represent the empty tree; 
we use a bud node, introduced in Section~\ref{sec:bud}, to express it.

Now the storage sizes of these three kinds of nodes are more balanced than the original definition:
$288$ bits for a leaf, $32 + 32 + 256 = 320$ bits for a branch,
and $k + 32$ bits for an extender.
This balanced-size design of nodes mitigates the problem of internal fragmentation.

\subsubsection{Model of Plebeia Trees}
\label{sec:modelOfPlebeiaTree}

A Plebeia tree $t$ represents a key--value store $\model{t}$, which we call the \emph{model} of $t$, defined as follows.
For a key $k$, $\model{t}(k)$ is the value $v$ if there is a path labeled by $k$ from the root of $t$ to a leaf \fstarinline{Leaf(v,h)}; if such a leaf does not exist, then $\model{t}(k)$ is undefined ($\model{t}(k) = \bot$).
We write $\DOM(t)$ for the domain of $\model{t}$.
Since only leaves can keep values, $\DOM(t)$ for an MPT $t$ satisfies the following \emph{prefix-freedom} property: for any distinct two keys in $\DOM(t)$, neither is a prefix of the other.

\subsubsection{Structural Invariants of Plebeia Trees}

% \begin{remark}
%   \label{rem:restrictionOnExtender}
Plebeia does not allow (1) an extender to be a direct subnode of another extender and (2) an extender to have the empty key.
These two restrictions are to ensure that there is a canonical representation of a key--value store by a Plebeia tree; if these restrictions were not met, then one could add arbitrarily many extenders below a branch without changing the key--value store it represents.
We name these invariants on the structure of a Plebeia tree (called \emph{structural invariants}) as follows.
\begin{description}
  \item[(SI1)]\label{inv:consecutiveExtender} An extender must not have another extender as its direct child.
  \item[(SI2)]\label{inv:noEmptyKey} An extender must not have an empty key.
\end{description}

\subsection{Functions for Manipulating Plebeia Trees}
\label{sec:manipulatingFunctions}

Plebeia provides the following tree-manipulating functions.
\begin{itemize}
  \begin{sloppypar}
  \item
    \fstarinline{get_value : t:node -> k:key -> option!\texttt{value}!}.
    It returns \fstarinline{Some v} where \fstarinline{v} is the value on the leaf reached by following the key \fstarinline{k} from the root of the node \fstarinline{t}; if such a leaf does not exist, it returns \fstarinline{None}.
  \end{sloppypar}
  \begin{sloppypar}
  \item
    \fstarinline{subtree : t:node -> k:key -> option node}.
    It returns \fstarinline{Some t'} if \fstarinline{t'} is the tree reached by following the path according to key \fstarinline{k} from the root of \fstarinline{t}; if such \fstarinline{t'} does not exist, it returns \fstarinline{None}.
  \end{sloppypar}
  \begin{sloppypar}
  \item
    \fstarinline{insert : t:node -> k:key -> !\texttt{v:value}! -> option node}.
    It returns \fstarinline{Some t'} if \fstarinline{t'} is the tree obtained by inserting a leaf with value \fstarinline{v} to \fstarinline{t} at the position reached by following the path represented by the non-empty \fstarinline{k} from the root of \fstarinline{t}; if such insertion would violate the prefix-freedom condition (Section \ref{sec:modelOfPlebeiaTree}), it returns \fstarinline{None}.
  \end{sloppypar}
  \begin{sloppypar}
  \item \fstarinline{update : t:node -> k:key ->!\texttt{v:value}! -> option node}.
    It returns \fstarinline{t'} that is obtained by replacing the leaf at the position reached by following the path represented by the key \fstarinline{k} from the root of \fstarinline{n} with a leaf with the value \fstarinline{v}; if such a leaf does not exist at the position represented by \fstarinline{k}, it returns \fstarinline{None}.
  \end{sloppypar}
  \begin{sloppypar}
  \item \fstarinline{delete : t:node -> k:key -> option node}.
    It returns  \fstarinline{Some t'} if \fstarinline{t'} is obtained by deleting the leaf at the position reached by following the path represented by \fstarinline{k} from the root of \fstarinline{t}; if such a leaf does not exist, it returns \fstarinline{None}.
  \end{sloppypar}
\end{itemize}

\subsection{Functional Requirements for the Tree-Manipulating Functions}
\label{sec:functionalCorrectness}

We state the functional requirements for the tree-manipulating functions.
In the following, we describe those for \fstarinline{get_value} and \fstarinline{insert}.

The expected property of \fstarinline{get_value} can be stated as follows:
for any Plebeia tree \fstarinline{t} and key \fstarinline{k},
\begin{itemize}
\item[(1)] $\mathfstar{get_value t k = Some(v)} \implies \model{t}(k) = v$ and
\item[(2)] $\mathfstar{get_value t k = None} \implies \model{t}(k) = \bot$,
\end{itemize}
where $\model{t}$ is the model of $t$ defined in Section~\ref{sec:modelOfPlebeiaTree}.
The above requirements express that the function call \fstarinline{get_value t k} should look up the value of $k$ in the mapping $\model{t}$.

The expected property of the \fstarinline{insert} is as follows:
For any Plebeia tree \fstarinline{t}, key \fstarinline{k}, and value \fstarinline{v},
\begin{itemize}
\item[(1)] $\mathfstar{insert t k v = Some t'} \implies$\\
  $\model{t}(k) = \bot \wedge \model{t'}(k) = v\ \wedge \forall k'.\ k' \neq k \implies \model{t}(k') = \model{t'}(k')$ and
  \item[(2)] $\mathfstar{insert t k v = None} \implies $\\
    $\{k\}\ \cup\ \DOM(t)\ \mathrm{is\ not\ prefix\ free}\ \vee\ k \in \DOM(t)$.
\end{itemize}
The first statement says that then the model of \fstarinline{t'} is the extension of $\model{t}$ with $\{k \mapsto v\}$ if \fstarinline{insert t k v} evaluates to \fstarinline{Some t'}.
The second statement says that, if \fstarinline{insert t k v} evaluates to \fstarinline{None} (hence fails), this means inserting \fstarinline{k} to $\DOM(n)$ violates the prefix-freedom condition or $k$ already exists in $\DOM(t)$.

\section{Extensions in the implementation}
\label{chap:extension}

In the actual implementation, Plebeia trees are implemented with several extensions for efficiency and persistence.
We describe these extensions in this section.

\subsection{Extension to Plebeia Trees}
\label{sec:extensionPlbeiaTrees}

The type of Plebeia trees with these extensions are as follows.
\begin{minted}[escapeinside=!!]{OCaml}
  type node = | Leaf of !\texttt{value}! * option index * option hash
              | Branch of node * node * option index * option hash
              | Extender of key * node * option index
              | Bud of option node * option index * option hash
              | Disk of index
\end{minted}
To explain this definition, we explain how it is obtained by gradually extending the original definition in Section~\ref{sec:typeForPlebeiaTrees}.

\subsubsection{Serialization of Trees}

Since a Plebeia tree is often too large to be stored on memory, it is defined so that a part of a tree can be serialized to external storage; hence the definition of \fstarinline{node} is extended to the following, in which the changes from the original definition are marked.
\begin{minted}[escapeinside=!!]{OCaml}
  type node = | Leaf of !\texttt{value}! * !\emphmarker{\texttt{index}}! * hash
              | Branch of node * node * !\emphmarker{\texttt{index}}! * hash
              | Extender of key * node * !\emphmarker{\texttt{index}}!
              | !\emphmarker{\texttt{Disk of index}}!
\end{minted}
We (1) add a constructor \fstarinline{Disk} and (2) attach fields of type \fstarinline{index} to each constructor.
The type \fstarinline{index} represents the addresses in the external storage.
\fstarinline{Disk i} expresses the node that is stored at the address \fstarinline{i} of the storage.
The \fstarinline{index} fields in the constructors \fstarinline{Leaf}, \fstarinline{Branch}, and \fstarinline{Extender} express to which address each node is supposed to be stored.

\subsubsection{Lazy Computation of Hashes and Indices}
\label{sec:lazyComputationOfHashesAndIndices}

The Merkle hash and the index of a node are not computed eagerly; they are computed only when the node is serialized to the storage.
This design is to prevent the computation of the Merkle hash and the index of an ephemeral node that is temporarily constructed on memory and hence is not needed to be serialized to storage.
To support such lazy computation of hashes and indices, we make the \fstarinline{index} and \fstarinline{hash} fields of \fstarinline{Leaf}, \fstarinline{Branch}, and \fstarinline{Extender} constructors to optional types:
\begin{minted}[escapeinside=!!]{OCaml}
  type node = | Leaf of !\texttt{value}! * !\emphmarker{\texttt{option}}! index * !\emphmarker{\texttt{option}}! hash
              | Branch of node * node * !\emphmarker{\texttt{option}}! index * !\emphmarker{\texttt{option}}! hash 
              | Extender of key * node * !\emphmarker{\texttt{option}}! index
              | Disk of index
\end{minted}
The \fstarinline{option index} and \fstarinline{option hash} fields are \fstarinline{None} until they are computed.
We say a node is \emph{indexed} if its field of type \fstarinline{index option} field is of the form \fstarinline{Some _}; a node is \emph{hashed} if (1) it has \fstarinline{hash option} field  which is of the form \fstarinline{Some _} or (2) it is of the form \fstarinline{Extender(_,n,_)} and \fstarinline{n} is hashed.

The computation of these fields happens according to the following rules:
\begin{itemize}
\item The hash and the index of a node are computed when the node is serialized to storage.
\item All the descendants of a node \fstarinline{n} have to be indexed before \fstarinline{n} is indexed.
\item All the descendants of a node \fstarinline{n} have to be hashed before \fstarinline{n} is hashed.
\end{itemize}
The second rule is to prevent a dangling pointer in the storage.
The third rule is due to the definition of a Merkle hash (Section~\ref{sec:mpt}).

\subsubsection{Bud Nodes}
\label{sec:bud}

Plebeia is being developed to implement the storage system of Tezos, which is represented as a file system with a directory structure, which requires Plebeia to implement a mechanism to represent the directory structure.
In order to represent a directory, a new kind of nodes called \emph{bud} is introduced in the Plebeia tree implementation.
The type \fstarinline{node} is extended as follows.
\begin{minted}[escapeinside=!!]{OCaml}
  type node = | Leaf of !\texttt{value}! * option index * option hash
              | Branch of node * node * option index * option hash
              | Extender of key * node * option index
              | !\emphmarker{\texttt{Bud of option node * option index * option hash}}!
              | Disk of index
\end{minted}
In the file system represented by this tree, a subtree whose root is a Bud node corresponds to a directory.
A node \fstarinline{Bud(optn,i,h)} has a subnode \fstarinline{optn} of \fstarinline{option node} type; this field is \fstarinline{None} if this bud node represents an empty directory.  Thus, the empty tree is represented by \fstarinline{Bud(None,i,h)}.

The model of a Plebeia tree with Bud nodes is extended from a key--value store to a \emph{nested} key--value store, corresponding to the fact that the Bud node represents a directory.
Concretely, the model of a Plebeia tree $t$ with bud nodes is defined as follows.
For a key $k$, $\model{t}(k)$ is:
\begin{itemize}
  \item The value $v$ if there is a path labeled by $k$ from the root of $t$ to a leaf \fstarinline{Leaf(v,h,i)};
  \item The nested key--value store $\model{t'}$ if there is a path labeled by $k$ from the root of $t$ to a bud \fstarinline{Bud(Some(t'),h,i)};
  \item The empty key--value store $\{\}$ if there is a path labeled by $k$ from the root of $t$ to a bud \fstarinline{Bud(None,h,i)}; and
  \item Undefined ($\bot$) if such a leaf or bud does not exist.
\end{itemize}
$\DOM(t)$ is defined in the same manner as an MPT,
and the \emph{prefix-free} property is similarly satisfied on the domain of a Plebeia tree.
For example, for the following tree (\fstarinline{index} and \fstarinline{hash} fields omitted for brevity):
\begin{minted}[]{ocaml}
  Branch(Branch(Leaf A,
                Bud(Some(Branch(Leaf B, Bud None)))),
         Extender([R;L],Leaf D))
\end{minted}
its model is as follows.
$ 
\model{t} = \{
  \mathrm{LL} \mapsto \mathrm{A},
  \mathrm{LR} \mapsto \{ \mathrm{L} \mapsto \mathrm{B}, \mathrm{R} \mapsto \{\}\},
  \mathrm{RRL} \mapsto \mathrm{D}
\} 
$.
Notice how buds are used to express the nesting structure of the key--value store.

\subsection{Zippers}
\label{sec:zipperExplanation}

One of the problems in the tree-manipulating functions in Section~\ref{sec:functionalCorrectness} is that each call to them has to traverse a tree from its root.
This deteriorates the performance of these functions if the tree is large.
The implementation of Plebeia prevents this problem by using \emph{zippers}~\cite{huet1997zipper}.

A zipper is a purely functional data structure to express pointers to internal nodes.
Intuitively, a zipper is a tree with a \emph{cursor} that points to a node in the tree.
The type for zippers is defined as follows.%
\footnote{This definition is slightly simplified from the actual formalization.  See Remark~\ref{sec:zipperDeviation}.}
  \begin{minted}[escapeinside=!!]{OCaml}
type path = 
  | Top 
  | Left of (* down to left from a Branch *)
      path * node * option index * option hash 
  | Right of (* down to right from a Branch *)
      node * path * option index * option hash
  | Extended of path * key * option index (* down from an Extender node *)
  | Budded of path * option index * option hash (* down from a Bud node *)
type zipper = path * node
type budzipper = path * n:node{Bud? n}
  \end{minted}
  A zipper is a pair of (1) a \fstarinline{path}, which describes how the node pointed to by the cursor can be reached from the root node, and (2) the subtree, whose root node is pointed to by the cursor.
  The intuition of the type \fstarinline{path} is as follows.
\begin{itemize}
  \item \fstarinline{Top} represents the empty path. 
  \item \begin{sloppypar}
  To explain the intuition of \fstarinline{Left(p,n,i,h)} (resp., \fstarinline{Right(n,p,i,h)}),
  let \fstarinline{n'} be the branch reached from the root by following the path \fstarinline{p};
  then, the path \fstarinline{Left(p,n,i,h)} (resp., \fstarinline{Right(n,p,i,h)}) represents
  the path from the root to the left (resp., right) child of \fstarinline{n'}.
  The index \fstarinline{i} and the hash \fstarinline{h} are those for \fstarinline{n'}.
  The tree \fstarinline{n} is the right (resp., left) subtree of \fstarinline{n'};
  we call this \fstarinline{n} an \emph{untracked tree} at \fstarinline{Left(p,n,i,h)}
  (resp., \fstarinline{Right(n,p,i,h)}).
  \end{sloppypar}
  \begin{sloppypar}
  \item Let \fstarinline{n'} be the extender reached from the root by following the path \fstarinline{p};
  then, the path \fstarinline{Extended(p,k,i)} represents the path from the root to the child of
  \fstarinline{n'} obtained by following the key \fstarinline{k}.  The index \fstarinline{i}
  is that for \fstarinline{n'}.
  \end{sloppypar}
  \begin{sloppypar}
  \item Let \fstarinline{n'} be the bud reached from the root by following the path \fstarinline{p};
  then, the path \fstarinline{Budded(p,i,h)} represents the path from the root to the child of
  \fstarinline{n'}.
  The index \fstarinline{i} and the hash \fstarinline{h} are those for \fstarinline{n'}.
  \end{sloppypar}
\end{itemize}
The type \fstarinline{budzipper} is for the zippers whose cursor points to a \fstarinline{Bud} node.
The predicate \fstarinline{Bud? n}, which is automatically generated by \fstar from the definition of the type \fstarinline{node}, holds if and only if \fstarinline{n} is of the shape \fstarinline{Bud _}.

Using a zipper, one can replace the subtree pointed to by the cursor with another one in constant time in a purely functional manner.
A cursor can be moved up and down in the tree (also in constant time).

\begin{wrapfigure}[12]{r}[10pt]{0.43\textwidth}
    \centering
    \includegraphics[scale=0.8]{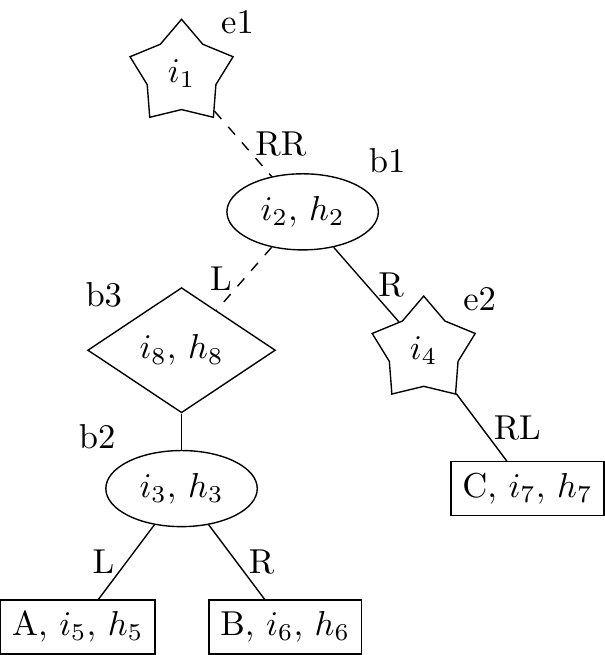}
    \caption{Illustration of \fstarinline{zipper_at_b3}.
    A star, an oval, a rectangle, and a diamond represent an extender, a branch, a leaf, and a bud, respectively.}
    \label{fig:zipperAtB3}
\end{wrapfigure}

For example, the following zipper:
\begin{minted}[escapeinside=!!]{OCaml}
let zipper_at_b3 =
  (Left(Extended(Top, [R;R], !$i_1$!),
        Extender([R;L],
          Leaf(!$C$!, !$i_7$!, !$h_7$!), !$i_4$!), !$i_2$!, !$h_2$!),
   Bud(Some(Branch(Leaf(!$A$!, !$i_5$!, !$h_5$!),
     Leaf(!$B$!, !$i_6$!, !$h_6$!), !$i_3$!, !$h_3$!)),!$i_8$!, !$h_8$!))
\end{minted}
represents the tree in Figure~\ref{fig:zipperAtB3} wherein the cursor points to the node \fstarinline{b3}
and the path from the root to \fstarinline{b3} is denoted by the dotted lines.
The second element of \fstarinline{zipper_at_b3} is the subtree rooted by \fstarinline{b3}.
The first element expresses the path from the root of the entire tree to the node \fstarinline{b3};
\fstarinline{b3} is reached from the extender at the root
following its sole edge labeled with \fstarinline{[R;R]} (reaching \fstarinline{b1})
and then the left branch of \fstarinline{b1}.

\subsection{Extensions to the Tree-Manipulating Functions}
\label{sec:extensionTreeManipulatingFunctions}

In the actual implementation, the tree-manipulating functions defined in Section~\ref{sec:manipulatingFunctions} are defined so that they manipulate a tree via a zipper whose cursor points to a \fstarinline{Bud} node.
For example, \fstarinline{insert} is defined as a function of type
\begin{minted}[escapeinside=!!]{OCaml}
  z:budzipper -> k:key ->!\texttt{v:value}! -> option budzipper
\end{minted}
instead of
\begin{minted}[escapeinside=!!]{OCaml}
  z:node -> k:key ->!\texttt{v:value}! -> option node
\end{minted}
The types of the other functions presented in Section~\ref{sec:manipulatingFunctions} are changed in a similar way.
This way, one can start a traversal of a tree from an internal node rather than the root, which reduces the cost of tree manipulations.
  
In addition to the functions in Section~\ref{sec:manipulatingFunctions}, Plebeia provides the following functions to serialize and deserialize a Plebeia tree.
\begin{itemize}
\item \fstarinline{commit_node : node -> node * index}, which writes
  each node of the tree recursively from bottom to top and returns a
  tree in which the index of each node is computed.  Plebeia uses a
  disk in an append-only style; therefore, \fstarinline{commit_node}
  has to store nodes at the indices which are disjoint from those
  already used.
\item \fstarinline{load_node : index -> node}, which reads data on the
  index from the disk and constructs a Plebeia tree recursively.  It
  does not always deserialize the whole tree; several subtrees are not
  read and left as a \fstarinline{Disk _} node.
\end{itemize}

\begin{remark}
  \label{sec:zipperDeviation}
  A tree-manipulation may lead to a recalculation of hashes and indexes.
  In the original implementation of Plebeia and our formalization,
  this recalculation is done lazily.
  To represent a hash and an index that need to be but have not been recalculated,
  the definition of \fstarinline{path} in the actual formalization is (an equivalent of) the following.
  \begin{minted}{ocaml}
type path = | Top 
  | Left  of path * node * option (option index * option hash)
  | Right of node * path * option (option index * option hash)
  | Extended of path * key * option (option index)
  | Budded of node * option (option index * option hash)
  \end{minted}
  \begin{sloppypar}
  Notice that (1) the field of type \fstarinline{option index * option hash} in
  \fstarinline{Left}, \fstarinline{Right}, and \fstarinline{Budded}
  is replaced with \fstarinline{option (option index * option hash)} and
  (2) the field of type \fstarinline{option index} in \fstarinline{Extended} is
  \fstarinline{option (option index)}.
  These fields are set to \fstarinline{None} if they need to be but yet to be recalculated.
  We omit this feature in this paper; see the source code
  for the actual formalization.
  \end{sloppypar}
\end{remark}

\subsection{Structural Invariants for Extended Plebeia Trees}
\label{sec:structuralInvariantsExtended}

The extensions on Plebeia trees presented in Section~\ref{sec:extensionPlbeiaTrees}, especially the lazy-computation feature in Section~\ref{sec:lazyComputationOfHashesAndIndices}, requires the following structural invariants.
\begin{description}
  \item[(SI3)]\label{inv:hashComputed} If a node is hashed, its direct child must be hashed.
  \item[(SI4)]\label{inv:indexedHashed} If a node is indexed, it must be hashed.
  \item[(SI5)]\label{inv:indexed} If a node is indexed, its direct child must be indexed.
\end{description}

There is an additional restriction on buds.
There should not be consecutive occurrences of buds; this forces a canonical structure of a Plebeia tree to express a key--value store.
A leaf should not be a direct child of a bud: this is to forbid the empty file name in the file system implemented with a Plebeia tree.
\begin{description}
\item[(SI6)]\label{inv:consecutiveBud} A bud must not have another bud or leaf as its direct child.
\end{description}

\subsection{Functional Requirements of Extended Tree-Manipulating Functions}
\label{sec:functionalCorrectnessExtended}

The functional requirements described in Section~\ref{sec:functionalCorrectness} are restated so that each tree-manipulating function can receive and return a zipper instead of a tree.
For this purpose, we need to define the model of a \fstarinline{budzipper}.

The model $\model{(p,n)}$ of a zipper $(p,n)$ is defined as a pair of the model of the path $\model{p}$ and that of the node $\model{n}$.
$\model{n}$ is the same as that in Section~\ref{sec:bud}.
$\model{p}$ is a list of pairs of a nested key--value store and a key; in addition to the key from the root of the entire tree to $n$, the model $\model{p}$ carries the information on the ``untracked'' trees.
For example, let $s$ be a nested key--value store $
s = \{
  \mathrm{bird} \mapsto \mathrm{A},
  \mathrm{mammal} \mapsto \{
      \mathrm{dog} \mapsto \{ \mathrm{beagle} \mapsto \mathrm{B} \}, \mathrm{cat} \mapsto \mathrm{C}\},
  \mathrm{fish} \mapsto \mathrm{D}
\}
$.
The zipper whose cursor points to the root of the tree corresponding to $s$ is $([],s)$.
The model of the zipper whose cursor steps two levels down along the keys ``mammal" and ``dog" is
$([
  (\mathrm{dog}, \{\mathrm{cat} \mapsto \mathrm{C}\});
  (\mathrm{mammal}, \{\mathrm{bird} \mapsto \mathrm{A},\mathrm{fish} \mapsto \mathrm{D}\})
],
\{ \mathrm{beagle} \mapsto \mathrm{B} \})$.
The first element of the model is the list of ``mammal'' and ``dog'', each of which is associated with the nested key--value store that is not tracked.

On top of the above definition of a zipper, we can redefine the functional requirements.
For example, the requirement for \fstarinline{insert} is as follows:
For any \fstarinline{budzipper} $z := (p,n)$, key \fstarinline{k}, and value \fstarinline{v},
\begin{itemize}
\item[(1)] $\mathfstar{insert (p,n) k v = Some (p',n')} \implies$\\
  $\model{p}=\model{p'} \wedge \model{n}(k) = \bot \wedge \model{n'}(k) = v\ \wedge \forall k'.\ k' \neq k \implies \model{n}(k') = \model{n'}(k')$ and
  \item[(2)] $\mathfstar{insert (p,n) k v = None} \implies $\\
  $\{k\}\ \cup\ \DOM(n)\ \mathrm{is\ not\ prefix\ free}\ \vee\ k \in \DOM(n)$.
\end{itemize}
The condition $\model{p}=\model{p'}$ states that the cursor stays at the same position after an insertion.
Notice that \fstarinline{insert} is to insert values into the outermost MPT pointed to by the cursor;
to insert into an inner MPT, one first has to move the zipper down to the corresponding bud.

\section{Verification of Plebeia with \fstar}
\label{chap:verification}

This section explains how we verified a core part of Plebeia by \fstar.
We remark that the \fstar code we present in this section is sometimes simplified
from the original version.
\ifanonymized
For the complete code, see the supplementary material.
\else
\fi

\subsection{Overview}

The properties we proved on the implementation of Plebeia are summarized as follows.
\begin{itemize}
\item The tree-manipulating functions in Section~\ref{sec:manipulatingFunctions} preserve the structural invariants in Section~\ref{sec:structuralInvariantsExtended};
\item The tree-manipulating functions in Section~\ref{sec:manipulatingFunctions} satisfy the functional requirements mentioned in Section~\ref{sec:functionalCorrectnessExtended};
\item The functions that serialize and deserialize a Plebeia tree work correctly; and
\item The Merkle hash defined on Plebeia trees is relatively collision-resistant (i.e., collision-free assuming the hash functions on the primitive data).
\end{itemize}

Currently, we have ported four core files of Plebeia into \fstar:
(1) \texttt{node\_type.ml} in which the structure of Plebeia tree is defined,
(2) \texttt{cursor.ml} in which the tree-manipulating functions are defined,%
\footnote{
  In the actual implementation, a \fstarinline{zipper} is called as a \fstarinline{cursor}.
}
(3) \texttt{node\_storage.ml} in which the functions for (de)serializing Plebeia trees,
and (4) \texttt{hash.ml} in which the Merkle hash functions for Plebeia trees are defined.
We started the verification of Plebeia from these four files because:
(1) they implement the core functionality of Plebeia and therefore their implementation is already stable, and
(2) the code in other files often uses the OCaml standard library that has not been ported to \fstar yet.

\begin{figure}[t]
\centering
\includegraphics[scale=0.7]{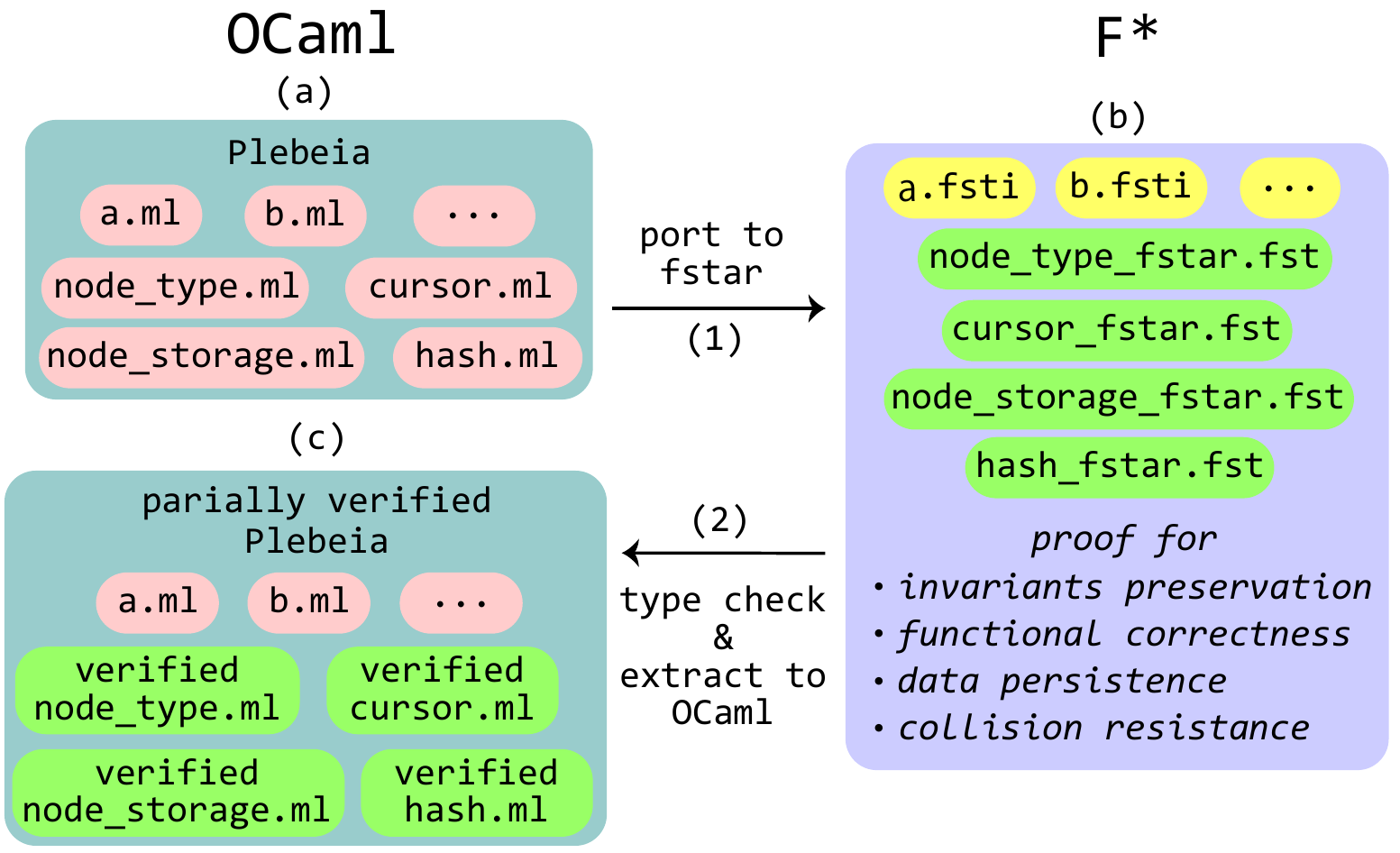}
\caption{How we verified the Plebeia library.}
\label{verify:howtoport}
\end{figure}

We conducted verification by the following workflow.
(The following explanation denotes parts in Figure~\ref{verify:howtoport} by the symbols in this figure.)
Plebeia (a) is implemented in OCaml.
It consists of several files, including \texttt{node\_type.ml}, \texttt{cursor.ml}, \texttt{node\_storage.ml}, and \texttt{hash.ml}.
As denoted by the arrow (1), we manually translated the programs in these files in (a) to
those of \fstar files (b).  These files include \fstar \emph{source files} whose extensions are \texttt{.fst}
(e.g., \texttt{node\_type\_fstar.fst},  \texttt{cursor\_fstar.fst}, \texttt{node\_storage\_fstar.fst}, and \texttt{hash\_fstar.fst} in (b)) and \fstar \emph{interface files} 
whose extensions are \texttt{.fsti}.
The source files include the definition of Plebeia functions, theorems stating correctness, and their proofs.
The interface files declare the types of the functions whose definitions have not been translated into \fstar yet; these types are treated as assumptions in the proofs
in \texttt{node\_type\_fstar.fst}, \texttt{cursor\_fstar.fst}, \texttt{node\_storage\_fstar.fst}, and \texttt{hash\_fstar.fst}.
From these ported \fstar files, we automatically extract OCaml source files (i.e., \texttt{node\_type.ml}, \texttt{cursor.ml}, \texttt{node\_storage.ml}, and \texttt{hash.ml}).
We replace \texttt{node\_type.ml}, \texttt{cursor.ml}, \texttt{node\_storage.ml}, and \texttt{hash.ml} in Plebeia with the extracted files and build the entire project.

One crucial merit of our workflow is that it enables \emph{gradual verification}
of the entire implementation.  One can run Plebeia \emph{before} the verification
of all the files completes.
Another benefit of using \fstar was that we were able to reuse most part of the implementation of Plebeia without rewriting them
because, as we mentioned in Section~\ref{chap:fstar},
the \fstar syntax is similar to that of OCaml.

\subsection{Verification of the Structural Invariants}
\label{sec:StructuralInvariants}

We explain how the verification of structural invariants of a Plebeia tree and a zipper proceeds in this section.

\subsubsection{Formalizing for the Structural Invariants for Plebeia trees}
\label{sec:fstarStructuralInvariantsPlebeiaTree}

We express the structural invariants of a Plebeia tree and a zipper
as \fstar functions whose return value is of type \fstarinline{bool} with effect \fstarinline{Tot}.
Concretely, the structural invariant for a Plebeia tree in Section~\ref{sec:structuralInvariantsExtended} is expressed by the
following function \fstarinline{node_invariant}.
\begin{minted}{ocaml}
let rec node_invariant (n:node) : Tot bool = 
  node_hashed_invariant n &&          (* SI3 *)
  node_index_and_hash_invariant n &&  (* SI4 *)
  node_indexed_invariant n &&         (* SI5 *)
  node_shape_invariant n &&           (* SI1, SI2, SI6 *)
  (match n with
  | Branch (l,r,i,h) -> 
      node_invariant l && node_invariant r
  ...)
\end{minted}
\begin{sloppypar}
The invariant is expressed as the conjunction of
\fstarinline{node_hashed_invariant n} that expresses the invariant on the hashedness of \fstarinline{v} (i.e., (SI3));
\fstarinline{node_index_and_hash_invariant n} that expresses the invariant on the dependency between indexedness and hashedness (i.e., (SI4));
\fstarinline{node_indexed_invariant n} that expresses the invariant on the indexedness of \fstarinline{v} (i.e., (SI5));
\fstarinline{node_shape_invariant n} that expresses the invariant on shape of \fstarinline{v} (i.e., (SI1), (SI2), and (SI6)); and
the condition to recursively check these invariants on the descendants of \fstarinline{v}.
A predicate \fstarinline{node_invariant (v:node)} evaluates to \fstarinline{true}
if the tree rooted by \fstarinline{v} satisfies the (SI1)--(SI6).
\end{sloppypar}

We only explain the implementation of \fstarinline{node_shape_invariant} here,
which is defined by the following function.
\begin{minted}[linenos,escapeinside=@@]{ocaml}
let node_shape_invariant (n:node) : Tot bool = match n with
  | Extender ([], _, _) -> false @\label{line:noemptypath}@   (* Checking SI2 *)
  (* Checking SI1 *)
  | Extender (_, (Extender _), _) -> false @\label{line:noconsecutiveextender}@
  | Extender (_, _, _) -> true
  (* Checking SI6 *)
  | Bud (Some (Bud _), _, _)    | Bud (Some (Leaf _), _, _)     -> false @\label{line:noconsecutivebudorleaf}@
  | Bud (Some (Branch _), _, _) | Bud (Some (Extender _), _, _) -> true
  | Bud (None, _, _) | Leaf _ | Branch _ -> true
\end{minted}
This predicate imposes the invariants (SI1), (SI2), and (SI6) in Section~\ref{chap:plebeiatrees}.
Line~\ref{line:noemptypath} imposes that there is no empty path originating from an extender;
Line~\ref{line:noconsecutiveextender} imposes that there are no consecutive extender nodes;
Line~\ref{line:noconsecutivebudorleaf} imposes that there is no bud node whose child is a bud or leaf node.

\subsubsection{Structural Invariants for Zippers}
\label{sec:fstarStructuralInvariantsZippers}

\begin{wrapfigure}[14]{r}{0.2\textwidth}
  % \centering
  \begin{center}
    \includegraphics[width=0.2\textwidth]{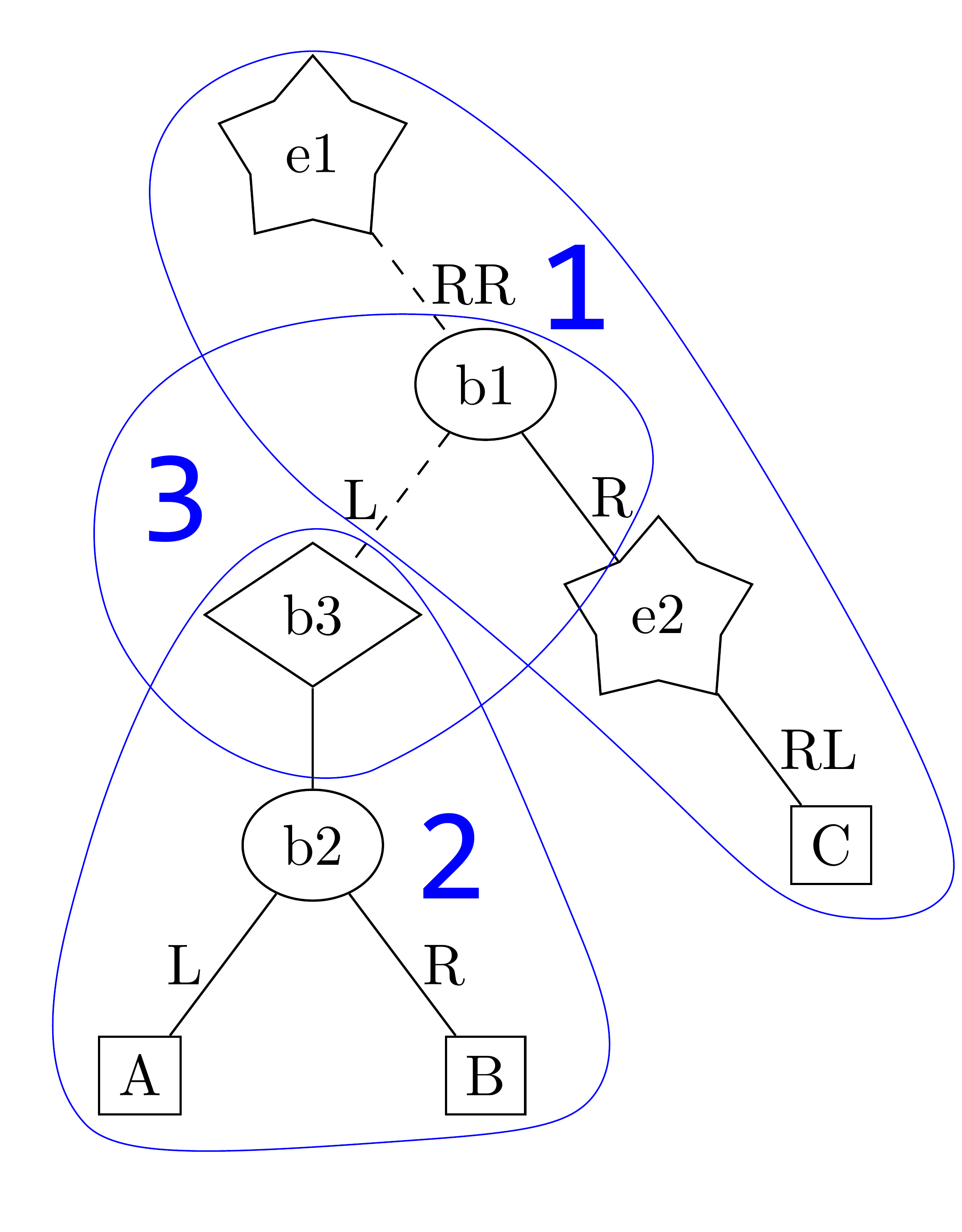}
  \end{center}
  \caption{
    Structural invariants for zippers.
  }
  \label{fig:zipper-invariants}
\end{wrapfigure}

To prove that the tree-manipulating functions in Section~\ref{sec:extensionTreeManipulatingFunctions} preserve the structural invariants (SI1)--(SI6), we formulate the invariants of a zipper that are preserved by these functions and that are strong enough to imply (SI1)--(SI6) for the entire tree to which the zipper is associated.
For notational convenience, we write \fstarinline{LR(p, n, i, h)} for a \fstarinline{path} value of the form \fstarinline{Left(p, n, i, h)} or \fstarinline{Right(n, p, i, h)} in the following.
Then, the invariants we require for a zipper to satisfy are stated as follows.
\begin{description}
  \item[(ZI-Node)] For a zipper \fstarinline{(_, n)}, the node \fstarinline{n} must satisfy (SI1)--(SI6).

  \item[(ZI-Path)] \begin{sloppypar}
  For a zipper \fstarinline{(p, _)}, if \fstarinline{p} is of the from
  \fstarinline{LR(p', n', i', h')}
  then (1-1) if \fstarinline{p'} is hashed,
  then \fstarinline{p} and \fstarinline{n'} must be hashed;
  (1-2) if \fstarinline{p'} is indexed,
  then \fstarinline{p} and \fstarinline{n'} must be indexed;
  (1-3) if \fstarinline{p} is indexed, then \fstarinline{p} must be hashed; and
  (1-4) \fstarinline{n'} must satisfy (SI1)--(SI6).
  For a zipper \fstarinline{(p,n)}, if \fstarinline{p} is of the form
  \fstarinline{Extended(p', k, i)}, then
  (2-1) if \fstarinline{p'} is indexed, then \fstarinline{p} must be indexed;
  (2-2) \fstarinline{p'} must not be the form of \fstarinline{Extended _}; and
  (2-3) \fstarinline{k} must be non-empty.
  For a zipper \fstarinline{(p,n)}, if \fstarinline{p} is of the form
  \fstarinline{Budded(p, k, i)}, then
  (3-1) if \fstarinline{p'} is hashed, then \fstarinline{p} must be hashed;
  (3-2) if \fstarinline{p'} is indexed, then \fstarinline{p} must be indexed; 
  (3-3) if \fstarinline{p} is indexed, then \fstarinline{p} must be hashed; and
  (3-4) \fstarinline{p'} must not be the form of \fstarinline{Budded _}.
  \end{sloppypar}
  \item[(ZI-Connect)] \begin{sloppypar}
  For a zipper \fstarinline{(p,n)},
  (1-1) if \fstarinline{p} is of the form \fstarinline{LR _} or \fstarinline{Bud _} and is hashed, then \fstarinline{n} must be hashed;
  (1-2) if \fstarinline{p} is of the form \fstarinline{Extended(p', _, _)} and \fstarinline{p'} is hashed, then \fstarinline{n} must be hashed;
  (2) if \fstarinline{p} is indexed, then \fstarinline{n} must be indexed; and
  (3) if \fstarinline{p} is of the form \fstarinline{Extended _}, then
  \fstarinline{n} must not be of the form \fstarinline{Extender _}.
  (4) if \fstarinline{p} is of the form \fstarinline{Budded _}, then
  \fstarinline{n} must not be of the form \fstarinline{Bud _} or \fstarinline{Leaf _}.
  \end{sloppypar}
\end{description}

Intuitively, for a zipper \fstarinline{(p,n)}, the property (ZI-Node) forces that
the node \fstarinline{n} satisfies (SI1)--(SI6).
The property (ZI-Path) guarantees that the part of the tree
expressed by the path \fstarinline{p} satisfies (SI1)--(SI6).
The property (ZI-Connect) guarantees that the connection between the path \fstarinline{p}
and the node of \fstarinline{n} does not violate (SI1)--(SI6);
for example, this property violates that \fstarinline{p} is of the form \fstarinline{Extended _}
and \fstarinline{n} is of the form \fstarinline{Extender _} at the same time
so that the tree to which the zipper \fstarinline{(p,n)} does not violate (SI4).

Figure~\ref{fig:zipper-invariants} explains these invariants schematically.
The structural invariants of the part 2 in the figure is established by (ZI-Node);
the part 1 is by (ZI-Path);
the part 3 is by (ZI-Connect).

\subsubsection{Proving the Preservation of the Structural Invariants}

Based on the definitions so far, we prove that the tree-manipulating functions
introduced in Section~\ref{sec:manipulatingFunctions} preserve the structural invariants for a zipper.
To state the preservation property compactly, we redefine the type of \fstarinline{zipper}
as follows.
\begin{minted}{ocaml}
type zipper = z:zipper{zipper_invariant z}
\end{minted}
The redefined type refines the original type for zippers
so that the inhabiting values satisfy the structural invariants for zippers.

This redefinition is inserted before the definition of the tree-manipulating functions.
Therefore, type-checking these functions proves the preservation of the structural invariants by the tree-manipulating functions.
For example, the type of function \fstarinline{insert} is stated as follows.
As mentioned in Section~\ref{sec:manipulatingFunctions},
a zipper that is given to or returned from some of the tree-manipulation functions
is restricted to the \fstarinline{budzipper}.
\begin{minted}[escapeinside=@@]{ocaml}
type budzipper = z:zipper{let _,n = z in Bud? n}
val insert: (z:budzipper) -> (k:key{k <> []}) -> @\texttt{value}@ -> option budzipper
\end{minted}
Proving \fstarinline{insert} has the above type constitutes the proof of
the preservation of the structural invariants by \fstarinline{insert}.

Quite surprisingly, this proof is done almost automatically:
\fstar backed by Z3 typechecks
\fstarinline{get_value}, \fstarinline{subtree},
\fstarinline{insert},
\fstarinline{create_subtree}, \fstarinline{update}, and \fstarinline{delete}
with only the following lemma on the length of a concatenated list:
\begin{minted}{ocaml}
  val list_app_length : (a:list 'a) -> (b:list 'a) ->
    Lemma (length (a @ b) = length a + length b)
\end{minted}
This lemma is used in typechecking an auxiliary function
that moves down the cursor of a zipper according to a key.

We also prove the following lemma that guarantees that 
the structural invariants for a zipper imply
(SI1)--(SI6) for the tree to which the zipper is associated.
\begin{minted}{ocaml}
val zipper_invariant_is_node_invariant : 
  z:zipper -> Lemma (requires True)
                    (ensures (let no = go_up_to_root z in node_invariant no))
                    (decreases (zipper_path_len z))
\end{minted}
\begin{sloppypar}
Notice that the type \fstarinline{zipper} of \fstarinline{z} works as the assumption
that \fstarinline{z} satisfies the structural invariants; therefore,
the \fstarinline{requires} clause requires just \fstarinline{True}.
The function \fstarinline{zipper_path_len z} computes the length of the path of \fstarinline{z}.
The clause \fstarinline{decreases (zipper_path_len z)} is required to conduct
the well-founded induction in the proof of this lemma (see below).
The function \fstarinline{go_up_to_root z} is the function that computes
the Plebeia tree to which \fstarinline{z} is associated.
\end{sloppypar}
\begin{minted}{ocaml}
val go_up_to_root : z:zipper -> Tot node (decreases (zipper_path_len z))
let rec go_up_to_root (p,n) = match p with
  | Top -> n | _ -> let Some z' = go_up (p,n) in go_up_to_root z'
\end{minted}
The type of \fstarinline{go_up}, which moves the cursor of a zipper up, is as follows:
\begin{minted}{ocaml}
val go_up :
  (c:zipper) -> Tot r:(option zipper{zipper_path_len r << zipper_path_len c})
\end{minted}
whose post-condition states that the length of the path of a returned zipper
is strictly shorter than that of the passed zipper.

Then, \fstarinline{zipper_invariant_is_node_invariant} can be proved as follows.
\begin{minted}{ocaml}
let rec zipper_invariant_is_node_invariant (p,n) = match p with
  | Top -> ()
  | _ -> let Some z' = go_up (p,n) in zipper_invariant_is_node_invariant z'
\end{minted}
The proof is essentially the %well-founded
induction on the length of the path \fstarinline{p}.
In the inductive step where \fstarinline{p} is not \fstarinline{Top},
\fstarinline{go_up (p,n)} generates the zipper \fstarinline{z'} in which its cursor is moved one-step up
from that of \fstarinline{(p,n)}.  Since the type of \fstarinline{go_up} guarantees that
the length of the path of \fstarinline{z'} is strictly shorter than that of \fstarinline{(p,n)},
we can use the induction hypothesis \fstarinline{zipper_invariant_is_node_invariant z'} here.
\footnote{Readers may be wondering why \fstar can see that \fstarinline{go_up (p,n)} never returns \fstarinline{None} when the type of \fstarinline{go_up} does not say anything about when \fstarinline{Some} is returned; it seems that \fstar combines the definition of \fstarinline{go_up} (not shown) and the fact that \fstarinline{(p,n)} is not \fstarinline{Top} to see it.}

\subsection{Verification of Functional Correctness}
\label{sec:FunctionalCorrectness}

\subsubsection{Definition of the Modeling Function}

Since the functional correctness is stated using the function $\model{\cdot}$,
we need to define a function to compute the model of a Plebeia tree.
First, we define the types of models of zippers, paths, and nodes as follows.
\begin{minted}[escapeinside=!!]{ocaml}
let is_model_of_node l =
  let keys = List.map fst l in is_prefix_free keys && is_sorted keys
type nestedvalue = Value of value | Map of model_of_node
and model_of_node = l:list (key * !\texttt{nestedvalue}!){ is_model_of_node l }
type model_of_path = list (key * model_of_node)
type model = model_of_path * model_of_node
\end{minted}
These definitions follow the definition of the model of a zipper described in Section~\ref{sec:functionalCorrectnessExtended}.
We encode a nested key--value store as an association list refined by a predicate \fstarinline{is_model_of_node}.
This predicate \fstarinline{is_model_of_node l} forces
(1) the keys in \fstarinline{l} to satisfy the prefix-freedom and
(2) the association list to be sorted in the lexicographic order over keys.
The former property is mentioned in Section~\ref{sec:modelOfPlebeiaTree}; the latter is for the convenience of the proof.

A function \fstarinline{modelize} that computes the model of a zipper
is defined as follows.
\begin{minted}{ocaml}
val modelize_node_aux : node -> model_of_node
let rec modelize_node_aux n = match n with
  | Leaf(v,_,_) -> [([],Value v)]
  | Branch(l,r,_,_) -> 
    let tl,tr = modelize_node_aux l, modelize_node_aux r in
      model_branch_lemma tl tr;
      (prepend_all [L] tl) @ (prepend_all [R] tr)
  | Extender(seg,n',_) -> prepend_all seg (modelize_node_aux n')
  | Bud(None,_,_) -> [([],Map [])]
  | Bud(Some n',_,_) -> [([],Map (modelize_node_aux n'))]

val modelize_node : 
  (n:node{match n with Bud _ -> true | _ -> false}) -> model_of_node
let modelize_node n =  
  let [([],Map res)] = modelize_node_aux n in res
val modelize_path : path -> model_of_path
let rec modelize_path p = match p with
  | Top -> []
  | _ -> 
      let p',n = go_up_to_bud_or_top (p,(Leaf(dummy,None,None))) in
      let n' = match n with Bud _ -> n | _ -> Bud(Some(n)) in
      let l = modelize_node n' in
      let k = key_to_nearest_bud p in
        (k, List.remove_assoc l k) :: modelize_path p'
val modelize : z:budzipper -> model
let modelize (p,n) = (modelize_path p,modelize_node n)
\end{minted}
\begin{sloppypar}
The function \fstarinline{modelize} computes the model
of a zipper \fstarinline{(p,n)} from
the model of \fstarinline{p} and
the model of \fstarinline{n}.
The function \fstarinline{modelize_node_aux} recursively constructs the model
following the definition of the model of a Plebeia tree discussed in Section~\ref{sec:bud}.
The function \fstarinline{prepend_all : k:key -> m:model_of_node -> model_of_node}
prepends \fstarinline{k} to every key of the model \fstarinline{m}.
The call to a lemma \fstarinline{model_branch_lemma tl tr} states that
the constructed model
\fstarinline{(prepend_all [L] tl) @ (prepend_all [R] tr)}
satisfies \fstarinline{is_model_of_node}.
\ifanonymized
For the definition and the proof of \fstarinline{model_branch_lemma},
see our formalization in the supplementary material.
\else
For the definition and the proof of \fstarinline{model_branch_lemma},
see the source code.
\fi
The function \fstarinline{modelize_path p} returns the model of path \fstarinline{p} as an association list following the definition described in Section~\ref{sec:functionalCorrectnessExtended}.
Given a path \fstarinline{p}, it creates a zipper whose path is \fstarinline{p} and whose cursor points to a dummy leaf node \fstarinline{Leaf(dummy,None,None)}.
Then, it moves up the zipper \fstarinline{(p,Leaf(dummy,None,None))} to \fstarinline{(p',n')} where \fstarinline{n'} is the nearest bud node; if there is no such bud node, then \fstarinline{n'} points to the root.
Using this zipper, it creates the model \fstarinline{l} of node \fstarinline{n'} (\fstarinline{modelize_node n'}) and computes the path from the root of \fstarinline{n'} to that of \fstarinline{n} (\fstarinline{key_to_nearest_bud p}).
Finally, it recursively computes \fstarinline{modelize_path p'} and combines with the entry for key \fstarinline{k}.
The value of \fstarinline{k} has to be the association list obtained by removing the entry for \fstarinline{k} from \fstarinline{l}, which is the dummy node created at the beginning of this function.
% \AI{No explanation about \fstarinline{modelize_path}.}
\end{sloppypar}

\subsubsection{Formalization of Functional Correctness}

Using the above modeling functions, we formalize the functional correctness of
\fstarinline{get_value}, \fstarinline{subtree}, \fstarinline{insert},
\fstarinline{create_subtree}, \fstarinline{update}, and \fstarinline{delete}.
For example, the functional correctness of \fstarinline{insert} is
formalized by the following declaration.
\begin{minted}[escapeinside=!!]{OCaml}
val insert_functionality :
    z:budzipper -> k:key -> v:!\texttt{value}! -> Lemma (
      let x = insert z k v in
      let p, l = modelize z in
      if is_prefix_free (List.Tot.map fst l) k then (
        match x with
        | Some z' -> let p', l' = modelize z' in
          p = p' !${\land}$! is_extended_key_value_store l' l k v
        | None -> False
      ) else (x = None))
\end{minted}
\begin{sloppypar}
The above code states the functional correctness of \fstarinline{insert}
discussed in Section~\ref{sec:functionalCorrectnessExtended}.
This code uses \fstarinline{is_extended_key_value_store l' l k v}, which holds if
(1) the mapping represented by \fstarinline{l'} is
an extension of the mapping represented by \fstarinline{l}
with $\{k \mapsto v\}$ and
(2) $k$ is not in the domain of the mapping represented by \fstarinline{l}.
\end{sloppypar}

\subsubsection{Proof of Functional Correctness}

We describe the strategy to prove the functional properties of
the tree-manipulating functions with \fstarinline{insert_functionality} as an example.
\ifanonymized
The correctness of the other functions is proved in the supplementary material.
\else
The correctness of the other functions is proved in the source code.
\fi

To prove \fstarinline{insert_functionality}, we designate
\begin{minted}[escapeinside=!!]{ocaml}
m_insert : (p,n):model -> k:key{k <> []} -> !\texttt{v:value}! -> model
\end{minted}
which expresses the desired behavior of \fstarinline{insert} on \emph{models of zippers}.
Then, we prove this theorem by reducing it to the functional correctness of \fstarinline{m_insert} as follows.
\begin{enumerate}
  \item[(A)]
  We prove the following equation on \fstarinline{insert} and \fstarinline{m_insert}:
  \[\fstarinline{modelize} \circ \fstarinline{insert} = \fstarinline{m_insert} \circ \fstarinline{modelize}.\]
  This proves that \fstarinline{insert} correctly reflects the behavior of
  \fstarinline{m_insert} with respect to the function \fstarinline{modelize}.

  \item[(B)] \begin{sloppypar}
  Then, we prove that \fstarinline{m_insert} is functionally correct in that
  it satisfies the following lemma \fstarinline{model_insert_functionality}:
    \begin{minted}[escapeinside=!!]{OCaml}
val model_insert_functionality :
  (p,n):model -> k:key{k <> []} -> v:!\texttt{value}! -> Lemma (
    let y = m_insert (p,n) k v in
    if is_prefix_free (List.Tot.map fst n) k then (
      match y with
      | Some (mp',ml') ->
        mp = mp' !${\land}$! is_extended_key_value_store ml' ml k v
      | None -> False
    ) else (y = None))
    \end{minted}
  \fstarinline{model_insert_functionality} states that \fstarinline{m_insert} satisfies
  the functional correctness of \fstarinline{insert} in Section~\ref{sec:functionalCorrectness}
  stated in terms of models of zippers.
  \fstarinline{model_insert_functionality} is easily proved by
  the induction on the structure of the association list \fstarinline{ml}.
  \end{sloppypar}
  
  \item[(C)] \begin{sloppypar}
  By using (A) and (B) in combination, we proved \fstarinline{insert_functionality}
  as follows.
  \begin{minted}[escapeinside=!!]{OCaml}
let insert_functionality z k v =
  insert_is_homomorphism z k v;
  model_insert_functionality (modelize z) k v
  \end{minted}
  \end{sloppypar}
\end{enumerate}

\subsection{Verification of Data-Persistence Process}
\label{sec:DataPersistence}

We verified the following three properties concerning \fstarinline{commit_node} and \fstarinline{load_node}, presented in Section~\ref{sec:extensionTreeManipulatingFunctions}, for the data-persistence process in Plebeia, 
\begin{description}
\item[(i)] \fstarinline{commit_node} stores a Plebeia tree in the correct format without overwriting existing serialized nodes.  (Recall that Plebeia uses a disk in an append-only style; see Section~\ref{sec:extensionTreeManipulatingFunctions}.)
\item[(ii)] From a disk in which Plebeia trees are stored in the correct format, \fstarinline{load_node} reads a well-formed Plebeia tree.
\item[(iii)] If a node is indexed with a valid index in a disk, then the indices of its descendants are valid in the disk.
\end{description}
To prove these properties, we modeled a disk and conducted structural induction on Plebeia trees.

One of the issues in the proof is that all functions used in proofs and predicates in \fstar must be pure and total, whereas functions that read/write from/to a disk implemented in Plebeia are neither pure nor total for the following reasons.
Plebeia accesses disk via memory-mapped file and treats it as a long byte array, hence not pure.
As for totality, in Plebeia, a function that reads/write a disk may not terminate if any condition are not assumed.
Indeed, \fstarinline{load_node} may diverge if they are called with a malformed disk, although such disk state does not occur as long as we access a disk only via correct disk reading/writing functions.
Even if we assume the well-formedness of a disk, proving the termination of disk-accessing functions is difficult because these functions lazily read a tree.

For the second problem of totality, we currently assume the termination of disk-accessing functions for a well-formed disk.
For the first problem about purity, We modeled a disk using \fstarinline{LowStar.Buffer.buffer}, which is the model of C arrays in \fstar introduced in Low*~\cite{lowstar}.
Using these models, we can model an effectful disk-accessing function as a pure function that passes the model that represents a disk state.
Then, we proved (1) the above properties for the modelized functions and (2) that the modelized functions are faithful to their effectful counterpart.
In the proof of the property (i), we also needed to prove that the set of indices modified by a call of \fstarinline{commit_node} is disjoint from those already used.
\fstarinline{LowStar.Buffer} provides various functions and propositions that are useful in the proof of this property.%
\footnote{The original \fstarinline{LowStar.Buffer} uses unsigned 32-bit integers as indices, whereas the maximum disk size supported by Plebeia is about $2^{37}$ bytes, which is not expressible in a 32-bit integer.
  For our proof, we implemented a variant of \fstarinline{LowStar.Buffer} that uses unsigned 64-bit integers as indices.}

To describe how the model of a disk looks like, consider the following example of the implementation of an effectful function \fstarinline{f} and its modelized version \fstarinline{spec_f}.
\begin{minted}{OCaml}
(* Original effectful disk-writing function, where storage is the type for
   memory-mapped disks.  Effect ST expresses that this function
   may read/write from/to memory. *)
let f (storage:storage) (i:index) (ch:char) : ST unit 
= (* Write two bytes to index i *)
  let buf = make_buf storage i in (* Create a buffer mapped to index i. *)
  set_char buf 10 ch;
  set_char buf 11 ch

(* The modelized version of f. *)
let spec_f
  (h:mem) (* Buffer modeled with LowStar.Buffer.buffer *)
  (storage:storage_model) (* Storage modeled with LowStar.Buffer.buffer *)
  (i:index) (ch:char)
  : Tot mem
= (* Relay heaps via arguments of functions *)
  let buf = spec_make_buf h storage i in (* Modelized make_buf *)
  let h1 = spec_set_char h  buf 10 ch in (* Modelized set_char *)
  let h2 = spec_set_char h1 buf 11 ch in
  h2 (* Explicitly return updated heap *)
\end{minted}
Then, in the type of \fstarinline{f}, we can state that \fstarinline{f} and \fstarinline{spec_f} are equivalent to each other.
\begin{minted}{OCaml}
(* Type declaration of function f *)
val f : storage:storage -> i:index -> ch:char -> ST unit
  (requires (fun _ -> true))
  (ensures (fun h _ h' ->
  (* In the postcondition of a function with ST effect,
     we can take h and h' that represent
     the state of the heap before and after calling this function. *)
     h' == spec_f h (model_of_storage storage) i ch
  ))
\end{minted}

To encode the assumption on the termination of the disk-accessing functions, we augmented their parameters with a natural number \fstarinline{fuel} and modified their implementation so that they decrement \fstarinline{fuel} by one at every recursive function call.
If the computation of a disk-accessing function terminates before \fstarinline{fuel} reaches $0$, then it returns \fstarinline{Some r} where \fstarinline{r} is the result; otherwise, it returns \fstarinline{None}.
The argument \fstarinline{fuel} is marked as a \emph{ghost} variable that does not appear in the extracted code.
We separately assumed an oracle \fstarinline{get_enough_fuel} that can compute a sufficiently large value for \fstarinline{fuel} from the state of a disk and a buffer.

\subsection{Verification of Relative Collision Resistance of the Merkle Hash of Plebeia tree}
\label{sec:hash}

% We also prove that the Merkle hash defined on the Plebeia tree is \emph{relatively collision-resistant}: the Merkle hash is collision-resistant \emph{under the assumption that} the blake2b hash function is cryptographically secure.
% %
% To describe the verification of the relative collision-resistance property, we first explain the definition of the collision-free property of hash functions.

% % In Plebeia, the Merkle hash of a Plebeia tree is used for the lightweight node of 
% % Tezos blockchain.
% % A lightweight node of the blockchain is used for a client of the blockchain network 
% % with limited memory size.
% % Compared to the full node, which holds entire blockchain status,
% % a lightweight node only holds the hash value of the Plebeia tree.
% % When a client retrieves the account data associated with an address from a full node,
% % it receives the account data together with a Merkle proof, from which 
% % the client node can verify that the received data is genuine.

% % The Merkle hash should be collision-free to prevent forging invalid account data.

% \subsubsection{Collision Free Hash Function}
% % For cryptographic hash functions, there are many desired security properties.
% % For example, the hash function should be preimage resistance\cite{}, which means that 
% % the original value should not be restored from a the value's hash.

We also prove that the Merkle hash defined on the Plebeia tree is \emph{relatively collision-resistant}: the Merkle hash is collision-resistant \emph{under the assumption that} the blake2b hash function is cryptographically secure.
To describe the verification of the relative collision-resistance property, we first explain the definition of the collision-free property of hash functions.

The collision-free property
\cite{DBLP:conf/eurocrypt/Damgard87} 
is an indispensable property for the hash functions used in the authentication.
% , is one of the indispensable properties of a hash function.
% The collision of a hash function is a pair of different values 
% that are hashed to the same value.
% The definition of a collision $(a,b)$
% is written as a mathematical expression 
% $a \neq b \wedge f(a) = f(b)$ where $f$ is the hash function.
A \emph{collision} of a hash function $f$ is a pair $(a,b)$ such that $a \neq b \wedge f(a) = f(b)$.
If a collision of the hash function of Plebeia tree is found in a reasonable amount of time,
it can be abused to forge invalid account data in the Tezos blockchain.
Therefore, it is important to verify 
the collision-free property on the hash Merkle of Plebeia tree.

The collision-free property is usually formalized on a
parameterized hash functions
to formally state the computational infeasibility for finding a collision
\cite{DBLP:conf/eurocrypt/Damgard87}.
In this formalization, the probability of
finding a collision is bounded from the above by the inverse of
any polynomial function of the parameter.
This formalization would be suitable for the theorem provers specialized for cryptographic verification. 
However, we do not use this formalization
% this definition is not suitable for the formalization in \fstar, 
because this definition requires probabilistic evaluation, which \fstar is not good at now.
% would require complicated proof in \fstar.
% Instead, we used the following "Human Ignorance" approach.

% this definition requires probabilistic evaluation,
% However, this definition is not suitable for the Merkle hash because 
% this definition requires probabilistic evaluation,
% which would make the proof
% this , which requires reasonings on the real numbers.
% The library on real numbers is hardly equipped with \fstar; thus, it would require a lot of code
% to formalize collision-free property.

Instead, we use the \emph{human ignorance} approach introduced by
Rogaway~\cite{humancollision}, which is a proof technique for
algorithms that use cryptographic primitives and is used in the
formalization of the collision-free property of the Merkle hash
implemented in HACL*, a verified library for cryptographic
primitives~\cite{zinzindohoue2017hacl}.  In this approach, one can
prove some cryptographic security of such an algorithm---the algorithm to
compute the Merkle hash of Plebeia in our case---with respect to the
cryptographic safety of the cryptographic
primitives---blake2b~\cite{Blake2} in our case---on which the
algorithm is based.  The proof is formalized by defining a reduction
function that receives a collision of the Merkle hash and generates a
collision or a preimage of a certain value of the blake2b, either of
which destroys the cryptographic security of the blake2b.

% Let $f_k : \{0,1\}^* \to Y_k$ be a family of hash functions. The parameter $k$ is said to be a parameter
% of the hash function. 
% A collision of a hash function $f$ is a pair $(a,b) \in X^2$ such that $a \neq b \wedge f(a) = f(b)$.

% Though the hash function defined on the Plebeia tree is 
% not parameterized. Therefore 
% Therefore, we verified that it is "hard to find"
% the collision of the hash function of Plebeia tree.
% This statement is formalized as follows, with informal definitions.
% \begin{theorem}
%   Let $h : Plebeia tree \to \{0,1\}^{28}$ be a hash function defined in Plebeia.
%  Then, it is    
% \end{theorem}
% The informal definition of the collision-free property is 
% The formal definition of the collision-free property should be defined on the 
% parameterized hash function. 

% The followings are the example of the security properties for the 
% hash function $f$. \ref{}
% \begin{itemize}
%   \item First Preimage Attack: 
%     Given $y$, find $x$ such that $f(x) = y$
%   \item Second Preimage Attack: 
%     Given $x_1$, find $x_2$ such that $x_1 \neq x_2 \wedge f(x_2) = f(x_1)$
%   \item Collision: 
%     Find $x_1, x_2$  such that $x_1 \neq x_2 \wedge f(x_2) = f(x_1)$
% \end{itemize}

% Usually, the resistance to first and second preimage attack is induced from 
% the resistance to the collision attack. \ref{}
% Therefore, we proved the resistance to the collision attack on the Merkle hash 
% of Plebeia tree.

% We proved that the Merkle hash defined on the Plebeia tree
% is collision resistance.

\subsubsection{The statement of the relative collision-resistance}
% We formalized the collision-free property of the Merkle hash based on the method introduced in
% \cite{humancollision}.
% In order to prove the collision-free property of this Merkle Hash, 
As mentioned above, the Merkle hash of Plebeia is constructed from the blake2b hash function,
which we assume to be secure~\cite{blake2issecure,blake2isindiff}.
More precisely, we assume the following two cryptographic-security properties of function $B'$, which takes the first 222 bits
of blake2b:\footnote{As we see, the length of a Merkle hash can vary (depending on the kind of a node) but it is often 28 bytes, which is equal to 224 bits, in the current implementation.  We will explain the other two bits below.}
% truncates
% because the last two bits of the blake2b hash function is truncated in the Merkle hash.
% % Therefore we assumed the cryptographic security of the following hash function $B'$; \\
% \begin{itemize}
% \item $B' : \{0,1\}^* \to \{0,1\}^{222},\ 
% B'(x) = \mathrm{the\ first\ 222\ bit\ of\ blake2B}(x)$
% \end{itemize}
% We assumed the following two assumptions.
\begin{assumption}
  \label{B'iscollisionfree}
  $B'$ is collision-free: It is hard to find 
  a pair $a ,b \in \{0,1\}^*$ such that $ a \neq b \wedge B'(a) = B'(b)$.
\end{assumption}
\begin{assumption}
  \label{B'ispreimageresistance}
  It is hard to find a string $z$ that satisfies $\mlq\underbrace{0\cdots0}_{222}\mrq$; i.e., a preimage of 
  the all-zero-bit string.
\end{assumption}
% and defined a reduction function which generates a counterexample of either of the assumptions
% from a collision of the Merkle hash.
As we will see soon, the second assumption is essential to prevent a collision on a \fstarinline{Bud(None,_,_)} node
and the first assumption is for the other kinds of nodes.
% The first assumption is originated from the collision-free property of the hash function $B'$, and 
% the second is from the preimage resistance of $B'$.
% The cryptographic security of the function $B'$ is inherited from the cryptographic security of the
% blake2b hash function.
% These assumptions are reasonable since
% blake2b is determined to use the prefix bytes of the raw hash value when truncating the hash,
% and such usage is tested to be secure~\cite{blake2issecure,blake2isindiff}.
% By the hardness of finding $X$ \AI{What is $X$?}, we mean that
% $X$ is not likely to be found within a reasonable time.
% If you think the phrase "it is hard to find X" seems to be ambiguous as the mathematical statement, 
% you can paraphrase it into "X is not likely to be found within a practical time".
Mathematically, there \emph{is} a collision of $B'$, 
as is easily shown by the pigeonhole principle.
However, a collision or a preimage of the all-zero-bit string has not been found yet, 
and it is unlikely to be found within a reasonable amount of time.
% These facts are essential for the practical use of hash functions.
% These facts are essential for the practical use of the hash functions such as $B'$ or the Merkle hash.

% A hash generated by the \fstarinline{blake2B_28} function 
% is the first 28 of the original blake2B function, therefore 
% overwriting last 2 bit of the \fstarinline{blake2B_28} function won't affect the 

% We assume that the blake2B hash function is cryptographic hash function, 
% so it has these three resistance properties mentioned above.
Based on these assumptions, what we would like to prove is the
following theorem about the Merkle hash $\mathit{MH}$:
\begin{theorem}
  \label{hashreduction}
  If the collision of the Merkle hash (i.e., a pair of 
  Merkle tree $(n_1,n_2)$ such that $n_1 \neq n_2 \wedge \mathit{MH}(n_1) = \mathit{MH}(n_2)$) is given,  
  then either of the following can be efficiently constructed from the collision:
  \begin{itemize}
    \item[(C1)] a pair of strings $(x_1,x_2)$ that satisfies $x_1 \neq x_2 \wedge B'(x_1) = B'(x_2)$; or
    \item[(C2)] a string $y$ that satisfies $B'(y) = \mlq0\mrq^{222}$.
  \end{itemize}
\end{theorem}
Our proof is not fully formal.  We formally show that a certain function
can construct (C1) or (C2) from a collision of the Merkle hash.  Since
the time complexity of the function is linear in the size of the
input, we know (C1) or (C2) can be \emph{efficiently} constructed but
we do not formalize the argument about complexity.

% (C1) corresponds to Assumption~\ref{B'iscollisionfree}, and 
% (C2) corresponds to Assumption~\ref{B'ispreimageresistance}.
% % (C1) is the collision of the $B'$, and (C2) is the preimage of the value $\mlq0\mrq^{222}$ by $B'$. 
% If (C1) or (C2) can be efficiently found, it violates the 
% cryptographic security assumptions of the hash function $B'$. 
% Together with Theorem\ref{hashreduction} with the Assumption\ref{B'issecure}, 
% we can induce that the Merkle hash is cryptographic hash function.
% The point of this formalization is the 
% efficiency of the reduction algorithm. 
% Without efficiency, this reduction wouldn't violate the "it is hard to find X" property.
% The efficiency of constructing an instance of (C1) or (C2) leads to
% the violation of the hard-to-find property.
% % "it is hard to find X" property.
% In the actual proof, the algorithm's time complexity is proportional to the size of the Merkle trees.  \AI{Did you formally prove anything about the complexity?}

\subsubsection{Definition of the Merkle Hash of Plebeia Tree}
We now describe the Merkle hash of Plebeia in more detail.
% h(Leaf(v)) = hash222_2(v,0b10)
% h(Internal(n1,n2)) = hash222_2(h(n1) || h(n2) || chr(len(h(n2))-28), 0b00)
% h(Extender(seg,n)) = h(n) || SE(seg)
% h(Bud None) = 0{224}
% h(Bud (Some n)) = hash222_2(n,0b11)
% than the section\ref{sec:mpt}.
The Merkle hash of Plebeia is defined as follows.\footnote{
  The actual implementation of the Merkle hash uses 
some stateful operations, such as updating a 
bytestring or loading a hash value from the disk for better
performance.  In what
follows, we ignore \fstarinline{Disk} for simplicity and present a purely functional definition
but the actual proof deals with stateful functions, using a device similar to
the one used for verifying persistence.}
\begin{minted}[escapeinside=!!]{OCaml}
(* returns the first 28 bytes of blake2b hash *)
val blake2B_28 : bytes -> b:bytes{length b = 28}

let hash_222_2 b v =
  update b 27 ((get (blake2B_28 b) 27 & 0xfc) | (int_to_byte v))
let (^^) = Bytes.append

let hash = function
  | Leaf (v,_,_) -> hash_222_2 v 0b10
  | Branch (l,r,_,_) ->
      let hl,hr = hash l,hash r in
      hash_222_2 (hl ^^ hr ^^ int_to_byte (Bytes.length hr))) 0b00
  | Extender (n',k,_,_) -> hash n' ^^ (key_to_bytes k)
  | Bud(None,_,_) -> Bytes.make 28 '!\textbackslash!0'
  | Bud(Some(n'),_,_) -> hash_222_2 (hash n') 0b11
\end{minted}
Here, the function \fstarinline{blake2B_28} is the blake2b hash
function with the hash size parameter is set to 28 (bytes);
\fstarinline{key_to_bytes} and \fstarinline{int_to_bytes} are (injective) functions to compute some
byte sequence from a list of keys and integers, respectively.

The size of a Merkle hash depends on the constructor of a tree.
The length of the Merkle hash of \fstarinline{Leaf}, \fstarinline{Bud}, and \fstarinline{Branch} nodes is 28 bytes,
but the length of the Merkle hash of \fstarinline{Extender} node is greater than 28 bytes.
% From the invariant (SI5), the key of an \fstarinline{Extender} node is
% not empty; therefore \AI{?  I thought it would depend on how you compute the hash of a key.}, it is ensured that 
% the hash of a node \fstarinline{Extender(n',k,_,_)} differs 
% from that of a node \fstarinline{n'}.
% Additionally, from the invariant (SI4), 
% two trees representing same key-value data 
We use the last 2 bits of a 28-byte hash to identify the kind of a node.
The function \fstarinline{hash_222_2} is used to compute blake2b of
the first argument and overwrite the last two bits of the 28 bytes
according to the second argument.  (An \fstarinline{Extender} does not
use this function because the length of its Merkle hash tells that it
is an \fstarinline{Extender}.)

% prevent collision.
% \AI{``We rewrite the last two bits of the bash according to the constructor by using \texttt{hash\_222\_2}''?  ``Set'' can mean write two 1's (just as ``reset'' can mean write two 0's').   I'm just curious to know why this function is called \texttt{222\_2}.  Ah, 28 bytes is equal to 224 bits!}
% If we did not set these bits, for example,
% the tree \fstarinline{Branch(Leaf("A"),Leaf("B"))} and 
% the tree \fstarinline{Leaf((blake2B "A")!\^{}\^{}!(blake2B "B"))} would produce 
% the same hash value  \fstarinline{blake2B((blake2B "A")!\^{}\^{}!(blake2B "B"))}.
% We overwrite the last 2 bits of the hash instead of appending the bits to 
% reduce the disk size.

\subsubsection{Formalization of the collision-free statement in \fstar}
We formalized Theorem~\ref{hashreduction} as the following function
\texttt{merkle\_hash\_collision2hash\_collision},
which returns either \fstarinline{Collision}(corresponding to C1) or 
\fstarinline{Preimage}(corresponding to C2) from a collision of the Merkle Hash.

\begin{minted}[escapeinside=!!]{OCaml}
let first_222_bit b = update b 27 (get b 27 & 0xfc)
let hash b = first_222_bit (blake2B_28 b)

type instance_of_attack =
  | Collision of (s1s2:(Seq.seq UInt8.t * Seq.seq UInt8.t){
      let s1,s2 = s1s2 in s1 <> s2 !$\wedge$! hash s1 = hash s2 })
  | Preimage of (s:Seq.seq UInt8.t{ hash s = Seq.create 28 0uy })

val merkle_hash_collision2hash_collision : 
  n1:node{node_invariant n1} -> n2:node{node_invariant n2} -> 
  Pure instance_of_attack
    (requires (n1 <> n2 !$\wedge$! merkle_hash n1 = merkle_hash n2))
    (ensures (fun _ -> True))
let rec merkle_hash_collision2hash_collision n1 n2 = (* ... *)
\end{minted}

% The function 
% \texttt{merkle\_hash\_collision2hash\_collision}
% computes the instance of the collision which breaks the 
% cryptographic security of the hash function $B'$
% from the collision instance of Merkle Hash.

% Thus, we defined the function \fstarinline{spec_merkle_hash}, 
% which is a purely functional implementation of the hash calculation function, and 
% proved the collision-free property according to the implementation.
% This approach is similar to the one which we used in verification of data-persistence process.

The computational complexity of 
\texttt{merkle\_hash\_collision2hash\_collision} is $O(N_1+N_2)$,
where $N_{i}$ is the size of the tree $n_{i}$.
Currently, there is no feature implemented in \fstar to 
describe the computational complexity of a function.
% , therefore
% we can't describe the computational complexity of 
% the function \texttt{merkle\_hash\_collision2hash\_collision}. 
However, the implementation of the function
is apparent from the source code, which is simple enough to examine its complexity.
We place the source code of its implementation in Appendix~\ref{appendix:collisionsource}.

% We also verified that the Merkle hash of Plebeia tree is collision-free.
% A hash function $f:X \to Y$ is said to be collision-free if
% it is hard to find a collision-pair $x_1, x_2 \in X$ where $x_1 <> x_2 \wedge f(x_1) = f(x_2)$.
% A cryptographic hash function need to be collision-free to avoid tampering.
% In the case of Plebeia, which is used in the storage system of 
% the Tezos, tampering of a tree can lead to tempering of the balance of accounts.

% We will prove the collision-free of Merkle Hash with the assumption 

% The formalization of collision-free property is based on the 
% formalization referred in \cite{}.
% In the paper, the collision-free 

% We are not able to mention about computational complexity with \fstar. 

\subsection{Result of the Verification}

Our proof on the structural invariants of Plebeia trees in Section~\ref{sec:StructuralInvariants} consists of 341 lines of \fstar code,
whereas the functional correctness in Section~\ref{sec:FunctionalCorrectness} consists of 4416 lines of \fstar code, 
the correctness of the data-persistence process in Section~\ref{sec:DataPersistence} consists of 3394 lines of \fstar code,
and the relative collision-resistant in Section~\ref{sec:hash} consists of 224 lines of \fstar code.
\ifanonymized
Our proof is included in the supplementary material.
\else
Our proof is included in the source code.
\fi

We executed the \fstar compiler in an environment with 1.1GHz CPU and 32GB RAM.
the version of the \fstar compiler was 0.9.7.0-alpha1; the version of the backend SMT solver Z3 was 4.8.5.
The time spent for type checking was 1722 seconds.
The memory space used during the type checking was 4400 megabytes.%
\footnote{\fstar compiler can cache verification results for future verification.
We turned off the caching in measuring efficiency.}

We built Plebeia by linking the OCaml code extracted from our proof with the unverified OCaml code.
We confirmed that the resulting executable passed all the tests bundled in the original Plebeia implementation.
The time spent to execute the Plebeia benchmark with the extracted code was 275 seconds, whereas 
the original Plebeia ran in 261 seconds.

It is worth mentioning that we found a bug that existed in an old version of Plebeia
through our verification: The structural invariant (SI4) was not preserved
in \fstarinline{delete}.  This bug remains in Plebeia from the commit 518f00db on April 9th, 2019
to the commit 12045fb1 on July 23rd, 2019.  Fortunately, this bug had been, in fact, fixed
before we discovered although we discovered this bug independently of the developers.

\section{Related Work}
\label{chap:relatedwork}

Project Everest~\cite{bhargavan2017everest} is a project for implementing TLS 1.3
\cite{rfc8446}---one of the essential components in HTTPS protocol---in \fstar 
to obtain a verified TLS library written in OCaml and C.
HACL*~\cite{zinzindohoue2017hacl} and
EverCrypt~\cite{protzenko2020evercrypt} are 
the libraries developed in Project Everest.
HACL* is a verified library for cryptographic primitives, which are used in
Mozilla Firefox\footnote{\url{https://blog.mozilla.org/security/2017/09/13/verified-cryptography-firefox-57/}} and Tezos.
HACL* is implemented with Low*~\cite{lowstar}, which is a part of \fstar that supports extraction to the C language.
The properties verified in HACL* includes memory safety, freedom from several side-channel attacks,
and functional correctness of each cryptographic primitive. 
EverCrypt is a cryptographic library, which uses HACL* and assembly code verified with
Vale~\cite{valeassemblylanguage}, which is an intermediate language that can express an assembly.  % assembly language annotated with Hoare-style conditions.
EverCrypt provides a verified implementation of Merkle trees;
the cryptographic primitives implemented in EverCrypt are used in these implementations.

Lochbihler and Maric~\cite{merkletreefunctor} formalized 
the authenticated data structures, which generalizes the Merkle hash
for various types of trees, as the datatype in Isabelle/HOL framework.
They formalized the Merkle tree construction step as the Merkle functor, which 
generates a Merkle tree from a tree.
They verified the correctness of abstracted algorithms of Merkle hash and Merkle proofs, 
whereas we directly verified a production-level implementation of the Plebeia tree.

Verification of complex data structures has been conducted by several authors.
For example, \citet{redblack} verified red-black trees by a dependently-typed ML-based language DML.
\citet{jahobverify} verified various linked data structures by the Jahob verification system.
\citet{malecha2010toward} verified B+ trees with Coq and used it to implement a verified relational database management system (RDBMS).
\citet{lammich2010isabelle} implemented a fast and verified data-structure library for Isabelle.
The GitHub repository of \fstar\footnote{
  \url{https://github.com/FStarLang/FStar/tree/master/examples}
} also includes verified implementation of various data structures (e.g., Merkle trees and red-black trees).

\citet{bhargavan2016formal} propose a framework to use \fstar for verifying Ethereum smart contracts.  
They propose (1) a certified translation from Solidity, a high-level programming language to write Ethereum smart contracts, 
to a subset of F* called Solidity* and (2) a certified translation from Ethereum bytecode to a subset of F* called EVM*.
The translated code can further be verified using F*.
Their framework is for the correctness verification of a smart contract,
whereas we verified the correctness of a core part of Tezos, which is an infrastructure on which a smart contract can be executed.

\section{Conclusion}
\label{chap:conclusion}

We verified the correctness of Plebeia, an implementation of a key--value store using an MPT-like data structure.
To this end, we ported the core part of Plebeia to \fstar.
The properties we verified are: (1) the structural invariants of a Plebeia tree are preserved by the tree-manipulating functions,
(2) the functional correctness of the tree-manipulating functions,
(3) the data-persistence process of Plebeia works correctly, and
(4) the relative collision resistance of the functions that compute the Merkle hash of Plebeia tree.
To the best of our knowledge, our work is the first one that formally verified a product-level implementation of an MPT-like data structure, which is heavily used in blockchain systems.

\iffull
\begin{acks}
This research is partially supported by JST CREST Grant Number JPMJCR2012 and MEXT/JSPS KAKENHI Grant Number JP19H04084.
\end{acks}

\else
\fi

% \pagelimitmarker{23}

\bibliographystyle{ACM-Reference-Format}
\bibliography{references}
\appendix
\section{Source Code of the Collision Reduction Function}\label{appendix:collisionsource}
% \begin{sloppypar}
  This source code is
  simplified from the original version by removing
  the fields for an index and a hash from each tree constructor.
%   a simplified version of the source code 
% listed in the following URL
% (\url{https://gitlab.com/dailambda/plebeia/-/blob/2f0f0695272349cb8a081367002ab0597e794d95/fstar/hash_collision_proof.fst}).
% In the following source code,
% each tree constructor is simplified by removing the fields for an index and a hash.
% \end{sloppypar}

% \begin{sloppypar}
This function recursively destructs a collision of the Merkle hash to 
generate an instance of \fstarinline{the_instance_of_attack},
which is either a collision or a preimage of the all-zero-bit string of blake2b.
% If two trees are constructed from the different constructors,
% then it directly generates a instance of 
% \fstarinline{the_instance_of_attack} without a recursive call. 
% If constructed from the same constructor,
% it directly generates a instance of \fstarinline{the_instance_of_attack}, or
% recursively c
The function \fstarinline{merkle_hash_collision2hash_collision n1 n2} 
is recursively called at most $\min(h_1,h_2)$ times,
where $h_1$ and $h_2$ are the heights of the trees \fstarinline{n1} and \fstarinline{n2},
respectively.
% This function is efficient enough to prove Theorem~\ref{hashreduction}.
% \end{sloppypar}
% Therefore the computational complexity of this function is
% the 
% \vspace{1cm}

\begin{minted}[escapeinside=!!]{OCaml}
  
let first_222_bit b = update b 27 (get b 27 & 0xfc)
let hash b = first_222_bit (blake2B_28 b)

(* before_hash_seq satisfies the following equation
   hash (before_hash_seq n) == 
   first_222_bit (model_of_merkle_hash n) *)
val before_hash_seq : 
  (n:node{match n with | Bud (Some _) | Branch _ | Leaf _ -> True 
                       | _ -> False}) ->
  Tot (Seq.seq UInt8.t)
let before_hash_seq = (* ... *)

type instance_of_attack =
  | Collision of (s1s2:(Seq.seq UInt8.t * Seq.seq UInt8.t){
      let s1,s2 = s1s2 in s1 <> s2 !$\wedge$! hash s1 = hash s2 })
  | Preimage of (s:Seq.seq UInt8.t{ hash s = Seq.create 28 0uy })

val merkle_hash_collision2hash_collision : 
  n1:node{node_invariant n1} -> n2:node{node_invariant n2} -> 
  Pure instance_of_attack
  (requires (n1 <> n2 !$\wedge$!
             model_of_merkle_hash n1 = model_of_merkle_hash n2))
  (ensures (fun _ -> True))

let rec merkle_hash_collision2hash_collision n1 n2 = 
  (* exclude impossible patterns. *)
  begin match n1,n2 with
  | Leaf _, Leaf _ | Bud(Some _), Bud(Some _) | Bud None,_ | _,Bud None
  | Branch _, Branch _ | Extender _, Extender _ -> ()
  | Extender _,_ | _,Extender _ -> 
      (* a pair of Extender and another type of node
          should not be a collision of the Merkle hash 
          because the length of the Merkle hash differs *)
      assert (Seq.length (model_of_merkle_hash n1) <>
              Seq.length (model_of_merkle_hash n2))
  | Bud(Some _), Branch _ | Branch _, Bud(Some _)
  | Leaf _, Bud(Some _) | Bud(Some _), Leaf _
  | Leaf _, Branch _ | Branch _, Leaf _ ->
      (* a pair of Extender and Bud(Some _)
          should not be a collision of the Merkle hash 
          because the last 2 bit of the Merkle hash differs *)
      assert (last_2_bit_of_hash (model_of_merkle_hash n1) <> 
              last_2_bit_of_hash (model_of_merkle_hash n2))
  end;
  match n1,n2 with
  | Bud(Some n1'),Bud(Some n2') when 
      (model_of_merkle_hash n1' = model_of_merkle_hash n2') -> 
      (* (n1',n2') is a collision of the Merkle hash *)
      merkle_hash_collision2hash_collision n1' n2'
  | Extender (k1,n1'), Extender (k2,n2') -> 
      assert (k1 = k2);
      (* (n1',n2') is a collision of the Merkle hash *)
      merkle_hash_collision2hash_collision n1' n2'
  | Branch(l1,r1), Branch(l2,r2) when (
      model_of_merkle_hash l1 = model_of_merkle_hash l2 && 
      model_of_merkle_hash r1 = model_of_merkle_hash r2) -> 
      if l1 <> l2 then
        (* (l1,l2) is a collision of the Merkle hash *)
        merkle_hash_collision2hash_collision l1 l2
      else (* r1 <> r2 *)
        (* (r1,r2) is a collision of the Merkle hash *)
        merkle_hash_collision2hash_collision r1 r2
  | Bud None,n | n,Bud None -> 
      let s = before_hash_seq n in
      (* s is a preimage of zeros *)
      Preimage_of_Zeros s
  | _ -> 
      let s1 = before_hash_seq n1 in
      let s2 = before_hash_seq n2 in
      (* (s1,s2) is a collision of B' *)
      Collision (s1,s2)
\end{minted}

\end{document}